\documentclass{aa}
\usepackage{graphicx}
\usepackage{txfonts}
\usepackage[round]{natbib}
\def\Msol{\ifmmode \mathrm{M}_{\sun}\else M$_{\sun}$\fi}
\def\Lsol{\ifmmode \mathrm{L}_{\sun}\else L$_{\sun}$\fi}
\def\Lya{\ifmmode \mathrm{Ly}{\alpha}\else Ly$\alpha$\fi}
\def\kms{\ifmmode \mathrm{km~s}^{-1}\else km~s$^{-1}$\fi}
\def\Oii{[O\,\textsc{ii}]}
\def\Oiii{[O\,\textsc{iii}]}
\def\vel{{\rm v}}
\def\spose#1{\hbox to 0pt{#1\hss}}
\def\lta{\mathrel{\spose{\lower 3pt\hbox{$\mathchar"218$}}
     \raise 2.0pt\hbox{$\mathchar"13C$}}}
\def\gta{\mathrel{\spose{\lower 3pt\hbox{$\mathchar"218$}}
     \raise 2.0pt\hbox{$\mathchar"13E$}}}

\begin{document}
\title{The Edge of the M87 Halo and the Kinematics of the Diffuse
  Light in the Virgo Cluster Core\footnote{Based on data collected
  with VLT Kueyen at Cerro Paranal, Chile, operated by ESO during
  observing run 076.B-0086(A)}}

\author{Michelle Doherty\inst{1}, Magda Arnaboldi\inst{2,3}, 
Payel Das\inst{4}, Ortwin Gerhard\inst{4}
\and \\
J. Alfonso L. Aguerri\inst{5}, Robin
Ciardullo\inst{6}, John J. Feldmeier\inst{7}, \\
Kenneth C. Freeman\inst{8}, George H. Jacoby\inst{9}, 
Giuseppe Murante\inst{3}}

\institute{European Southern Observatory, Santiago, Chile;
  mdoherty@eso.org
\and European Southern Observatory, Garching, Germany; marnabol@eso.org 
\and INAF, Osservatorio Astronomico di Pino Torinese, 
     Pino Torinese, Italy; murante@oato.inaf.it
\and Max-Planck-Institut f\"{u}r extraterrestrische Physik, Garching,
     Germany; pdas@mpe.mpg.de, gerhard@mpe.mpg.de 
\and Instituto de Astrofisica de Canarias, Tenerife, Spain;
     jalfonso@ll.iac.es
\and Dept. of Astronomy and Astrophysics, Pennsylvania State
     University, University Park, PA, USA; rbc@astro.psu.edu
\and Dept. of Physics and Astronomy, Youngstown State University,
     Youngstown, OH, USA; jjfeldmeier@ysu.edu
\and Mount Stromlo Observatory, Research School of Astronomy and
     Astrophysics, ACT, Australia; kcf@mso.anu.edu.au
\and WIYN Observatory, Tucson, AZ, USA; jacoby@noao.org
}

   \authorrunning{M. Doherty et al.}
   \titlerunning{The M87 Halo and the Diffuse Light in the Virgo Core}
   
   \date{Received 16.12.08; revised 25.03.09; accepted 22.04.09}

 
  \abstract
   {}
   {To study the kinematics and dynamics of the extreme outer halo of
   M87, the central galaxy in the Virgo cluster, and its transition to
   the intracluster light (ICL).}  
 {We present high resolution FLAMES/VLT spectroscopy of intracluster
   planetary nebula (PN) candidates, targeting three new fields in the
   Virgo cluster core with surface brightness down to $\mu_B=28.5$.
   Based on the projected phase space information (sky positions and
   line-of-sight velocities) we separate galaxy and cluster components
   in the confirmed PN sample.  We then use the spherical Jeans
   equation and the total gravitational potential as traced by the
   X-ray emission to derive the orbital distribution in the outer
   stellar halo of M87. We determine the luminosity-specific PN number
   for the M87 halo and the ICL from the photometric PN catalogs and
   sampled luminosities, and discuss the origin of the ICL in Virgo
   based on its measured PN velocities.}
 {We confirm a further 12 PNs in Virgo, five of which are bound to the
   halo of M87, and the remainder are true intracluster planetary
   nebulas (ICPNs). The M87 PNs are confined to the extended stellar
   envelope of M87, within a projected radius of $\sim160$ kpc, while
   the ICL PNs are scattered across the whole surveyed region between
   M87 and M86, supporting a truncation of M87's luminous outer halo
   at a $2\sigma$ level. The line-of-sight velocity distribution of
   the M87 PNs at projected radii of 60 kpc and 144 kpc {{ shows
     (i) no evidence for rotation of the halo along the photometric
     major axis, and (ii) that the velocity dispersion decreases in
   the outer halo, down to $\sigma_{last}= 78\pm 25$ km~s$^{-1}$ at
   144 kpc.}}  The Jeans model for the M87 halo stars fits the observed
   line-of-sight velocity dispersion profile only if the stellar
   orbits are strongly radially anisotropic ($\beta\simeq0.4$ at
   $r\simeq 10$ kpc increasing to $0.8$ at the outer edge), and if
   additionally the stellar halo is truncated at $\simeq 150$ kpc
   average elliptical radius.  The $\alpha$-parameters for the M87
   halo and the ICL are in the range of values observed for old ($>
   10$ Gyr) stellar populations.}
 {Both the spatial segregation of the PNs at the systemic velocity of
   M87 and the dynamical model support that the stellar halo of M87
   ends at $\sim 150$ kpc.  We discuss several possible explanations
   for the origin of this truncation but are unable to discriminate
   between them: {{ tidal truncation following an earlier encounter of
   M87 with another mass concentration in the Virgo core, possibly
   around M84, early AGN feedback effects, and adiabatic contraction
   due to the cluster dark matter collapsing onto M87.}}  From the
   spatial and velocity distribution of the ICPNs we infer that M87
   and M86 are falling towards each other and that we may be observing
   them just before the first close pass. The new PN data support the
   view that the core of the Virgo cluster is not yet virialized but
   is in an ongoing state of assembly, and that massive elliptical
   galaxies are important contributors to the ICL in the Virgo
   cluster. }

\keywords{galaxies: clusters: individual (Virgo) --- stellar dynamics
--- (ISM:) planetary nebulae: general --- galaxies: halos ---
galaxies: elliptical and lenticular, cD --- galaxies: formation}

   \maketitle
%

\section{Introduction}

Over the past few years the diffuse intracluster light (ICL) has been
the focus of many studies, both in nearby \citep{fmm+04,mhfm05} and in
intermediate redshift clusters \citep{zwsb05,krick07}. It has been
found that the ICL is centrally concentrated and in many cases,
including the diffuse outer halos of galaxies, comprises $\sim$10\% of
the total starlight in the cluster \citep{zwsb05}, and up to as much
as $\sim35\%$ \citep{gzz07}.

Theoretical studies of the diffuse cluster light through simulations
predict that the ICL is unmixed and therefore should exhibit a fair
amount of sub-structure \citep{npa+03,mag+04,rmm06}. An important
contribution to the diffuse light in clusters may come from the
extended halos of giant galaxies: numerical simulations predict the
presence of such halos around isolated galaxies out to several hundred
kpc, consisting of stars shed by merging sub-units \citep{ans06}. When
these galaxies enter the cluster core, their halos would be stripped
first by the tidal fields and later by the tidal shocking in the
interaction with the cluster's central core and cD galaxy
\citep{rmm06,mgga+07}.

Indeed the deep image of the Virgo cluster core by \citet{mhfm05},
reaching $\mu_V = 28~$mag arcsec$^{-2}$, shows a variety of features
such as streamers, arcs and smaller features associated with
individual galaxies. It also shows faint, very extended diffuse halos
surrounding the large galaxies. In particular, around the giant
elliptical galaxy M87, the \citet{mhfm05} photometry reveals an
extended stellar envelope at very low surface brightness levels,
$\mu_V > 26.5~$mag arcsec$^{-2}$, with flattened isophotes
\citep[noted previously by][]{arp71, wbm97}, and out to $\sim 37'$
($\simeq 161$ kpc) along the semi-major axis.

The Virgo cluster has long been known to be dynamically unmixed, with
complex sub-structures.  This was first realized from the spatial and
velocity distribution of Virgo galaxies \citep[e.g.][]{bts87,bpt93}.
In particular Binggeli et al. (1993) found tentative evidence from the
asymmetry in the velocity distribution of dwarf spheroidal galaxies
that even the core of Virgo is not virialised, and suggested that the
cluster is dynamically young, with two sub-clumps M87 and M86 falling
in towards each other in the centre.

{ From photometry in $z\!\sim\!0.1$ clusters
  \citep{gonzalez05,krick07} and from kinematic studies of the ICL in
  nearby clusters \citep{aga+04,ga+05,ga+07} we have learned that the
  genuine ICL component, defined as the light radiated by stars
  floating freely in the cluster potential, and the extended halos of
  bright (elliptical) galaxies often overlap spatially, and cannot
  easily be distinguished from broad-band photometry alone. Kinematic
  information can complement the photometry. For surface brightness
  $\mu_B\lta25$mag/arcsec$^2$, integrated light spectroscopy can be
  used to measure the mean velocity and velocity dispersion in the
  outer halos of the brightest cluster galaxies \citep{st96,kelson02};
  however, reaching the faint surface brightness level of the true ICL
  component with this technique is very difficult.  Since planetary
  nebulae (PNs) follow light \citep[e.g.][]{coccato+09}, the
  spectroscopic study of these tracers, both in the extended halos and
  the ICL, offers a way to identify and measure the kinematics of
  these diffuse stellar components down to very faint surface
  brightness ($\mu_B<28.5$ in Virgo), but it is currently limited to
  clusters with distances $<100$ Mpc \citep{ga+05}.

For the Virgo cluster, there has been considerable success with a
two-step approach of identifying PN candidates with narrow-band
imaging followed by multi-object spectroscopy. }
\citet{afm+96} observed the outer regions of the giant
elliptical M86, measuring velocities for 19 objects. Three of these
turned out to be true ICPNs, with velocities similar to that of the
mean velocity of the Virgo cluster. Subsequently, 23 PNs
were detected in a spectroscopic survey with 2dF on the 4m
Anglo-Australian Telescope (Freeman et al. 2000\nocite{fac+00};
Arnaboldi et al. 2002\nocite{aan+02}). These results were all based on
single line identifications, although the second oxygen line was seen
with the right ratio in the composite spectrum of 23 PNs observed by
Freeman et al. (2000). The first confirmation based on detecting the
[OIII] doublet in a single PN spectrum was made in \citet{afo+03}.
Expanding on this early work, we began a campaign to systematically
survey PN candidates in the Virgo cluster using multi-object
spectroscopy with the FLAMES/GIRAFFE spectrograph on the VLT
\citep[hereafter A04]{aga+04}. A04 presented the first measurements of
the velocity distribution of PNs from three survey fields in the Virgo
cluster core and concluded that in two of these fields the light is
dominated by the extended halos of the nearby giant elliptical
galaxies, while the ICL component dominates the diffuse light in only
one field, where a `broad' line-of-sight velocity distribution is
measured, and all PNs are true `ICPNs'.

We here present PN velocity measurements from a further three
pointings in the heart of the cluster core. We emphasize that these
pointings are targeting faint surface brightness regions well outside
of individual galaxies, in order to trace the ICPNs expected to be
moving freely in the cluster potential, and thus to investigate the
dynamical state of the ICL and of the core of the Virgo cluster. The
photometric/geometric classification of PNs as ICPNs is in fact
revised later in the paper according to the dynamical information
obtained from the line-of-sight (LOS) velocities of the confirmed PNs.

In this paper we give a summary of our observations and data reduction
in \S~\ref{obs} { where we also discuss the sample completeness and
  show the final emission spectra.  The distribution of measured
  line-of-sight velocities (LOSVD) and the projected phase-space
  diagram for these PNs are presented in \S~\ref{LOSVD}. From these
  data we distinguish between ICPNs and PNs bound to the halo of M87.
  In \S~\ref{M87} we discuss the rotation, velocity dispersion and
  physical extent of the stellar halo of M87, using the velocities of
  the PNs bound to M87 in the combined data sets of this paper and A04.
} We then construct a dynamical model based on the gravitational
potential obtained from X-ray observations and the combined
absorption-line and PN velocity dispersion data for the galaxy.
Possible mechanisms for the truncation of M87's stellar halo are
discussed in \S~\ref{origint}. In \S~\ref{alphap} we compute the
luminosity-specific PN number $\alpha_{2.5}$ for both the M87 halo and
the ICL in Virgo, and in \S~\ref{VC} we discuss the implications of
the ICPN LOSVD for our understanding of the dynamical status of the
cluster core and the origin of the ICL in Virgo. Summary and
conclusions of the paper are given in \S~\ref{last}.

In what follows, we adopt a distance of $15$ Mpc for M87,
equivalent to a distance modulus of 30.88; therefore $1'' = 73$ pc.

\section{Observations\label{obs}}

The observations were taken in service mode (22 hrs, 076.B-0086 PI: M.
Arnaboldi) over the nights 25-28th March 2006 using the FLAMES
spectrograph on UT2/VLT in MEDUSA mode { which allows spectra to be
  taken through up to 132 fibers simultaneously}\footnote{See
  http://www.eso.org/sci/facilities/paranal/instruments/flames/
  overview.html.}.  The data were taken in clear conditions with
seeing $<0.9''$. We used the high resolution grism HR504.8 centred at
504.8nm and with wavelength coverage 250\AA\ and spectral resolution
20,000. With this setup, the instrumental broadening of the arc lines
is FWHM = 0.29 \AA\ or 17 kms$^{-1}$, and the error on the wavelength
measurements is 0.0025 \AA\ or 150 ms$^{-1}$ \citep{Royer+02}.

Figure~1 shows the location of the selected fields
targeted with FLAMES, including the three previous fields FCJ, CORE
and SUB presented in A04, and the three new fields F4, F7\_1, F7\_2.
The photometry used for the selection of PN candidates is from
\citet{fcjd03}; his fields F4 and F7 contain the FLAMES fields F4 
and (F7\_1, F7\_2), respectively.

\begin{figure*}
\begin{centering}
\includegraphics[scale=0.7]{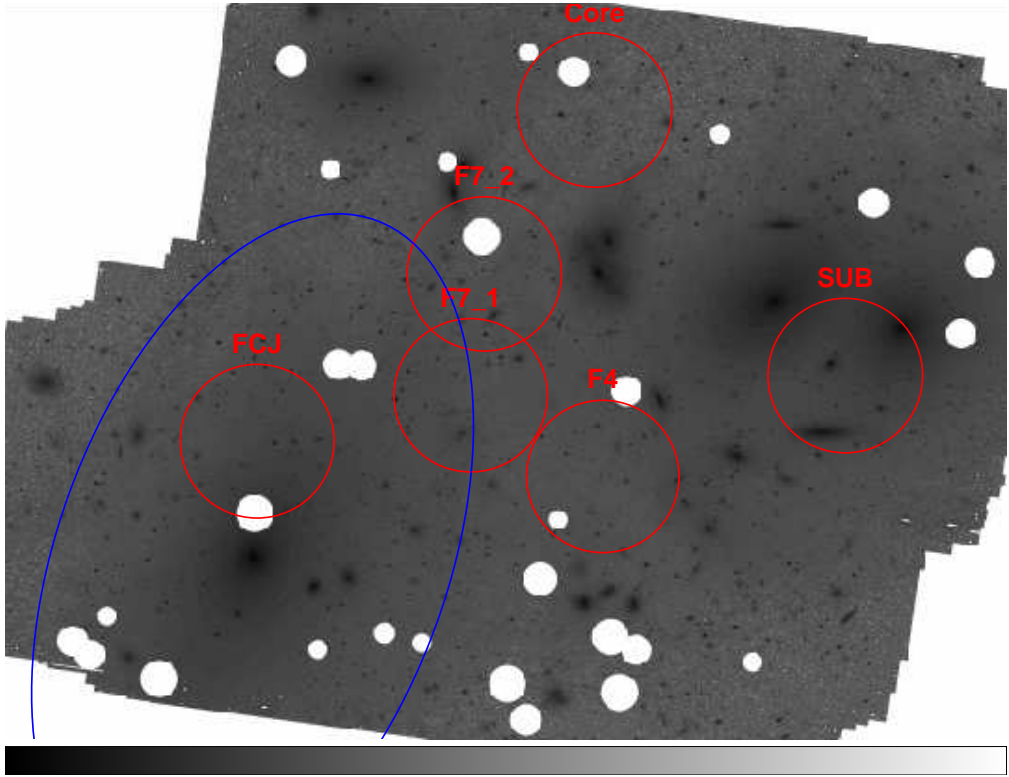}
\caption{Deep image of the Virgo cluster core showing the
diffuse light distribution \citep{mhfm05}, with  
our target fields superposed. Target fields of the previous
spectroscopy (A04) are shown as red circles and our new target fields
as well. The blue ellipse shows the boundary used in the dynamical
modeling in \S~~\ref{sec:dynamics}.}
\end{centering}
\label{fig:ourfields}
\end{figure*}

\subsection{Data reduction and sample completeness\label{datareduction}}

The data were reduced using the GIRAFFE pipeline\footnote{The GIRAFFE
  pipeline is available at\\ http://girbldrs.sourceforge.net.}
including bias subtraction, determining fiber location on the CCD,
wavelength calibration, geometric distortion corrections and
extraction of the one-dimensional spectra. The co-addition was carried
out separately as a final step on the one-dimensional spectra as the
fibers are allocated in a slightly different order for MEDUSA plates 1
and 2 and the pipeline does not account for this.

\begin{table}[h]
\caption{Observed Fields and Spectroscopic confirmations}
\begin{tabular}{cccc}
\hline 
Field  &  F7\_1       &  F7\_2       &  F4      \\  
$\alpha$(J2000.0) &  12 28 53.70 &  12 28 46.40 &  12 27 43.35 \\ 
$\delta$(J2000.0) & +12 44 32.5  & +13 00 20.5  & +12 33 57.5  \\ 
N$_{tot}^a$           & 22           &  28          &  25          \\ 
N$_{cmp}^b$           & 13           &  12          &   8            \\
N$_{em}^c$          & 9            &  5           &   5          \\ 
N$_{PN}^d$          & 6            &  4$^e$       &   3          \\ 
\hline\\

\end{tabular} \\
$^a$ Total number of targets with allocated fibers.\\
$^b$ Number of targets with allocated fibers and $m_{5007}$ brighter
or equal to the completeness magnitude limit of the photometric
survey.\\
$^c$ Number of spectra with detected emission { line}.\\
$^d$ Number of spectra with both \Oiii$\lambda\, 4959,5007$ \AA\ detected.\\
$^e$ One of these is in common with F7\_1, i.e., we have only 12
confirmed PNs in total, one of which was observed twice.\\
\label{tab:rates}
\normalsize
\end{table}

\begin{figure}
\includegraphics[scale=0.35,angle=90]{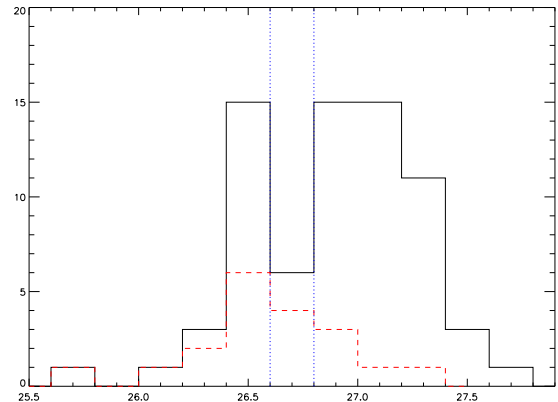}
\caption{Histogram showing the $m_{5007}$ magnitudes of all our
observed targets (solid black line) over-plotted with those where
emission lines were detected (red dashed line). The blue dotted lines
show the photometric completeness limits for target fields F4 (26.6)
and F7 (26.8).}
\label{fig:completeness}
\end{figure}

Table~\ref{tab:rates} shows the number of spectroscopically confirmed
emission-line objects and planetary nebulas, with respect to the
number of candidates targeted ($N_{tot}$), and the number of {
  candidates targeted} above the photometric completeness limit for
each field ($N_{cmp}$).

\begin{figure*}
\includegraphics[angle=90,scale=0.8]{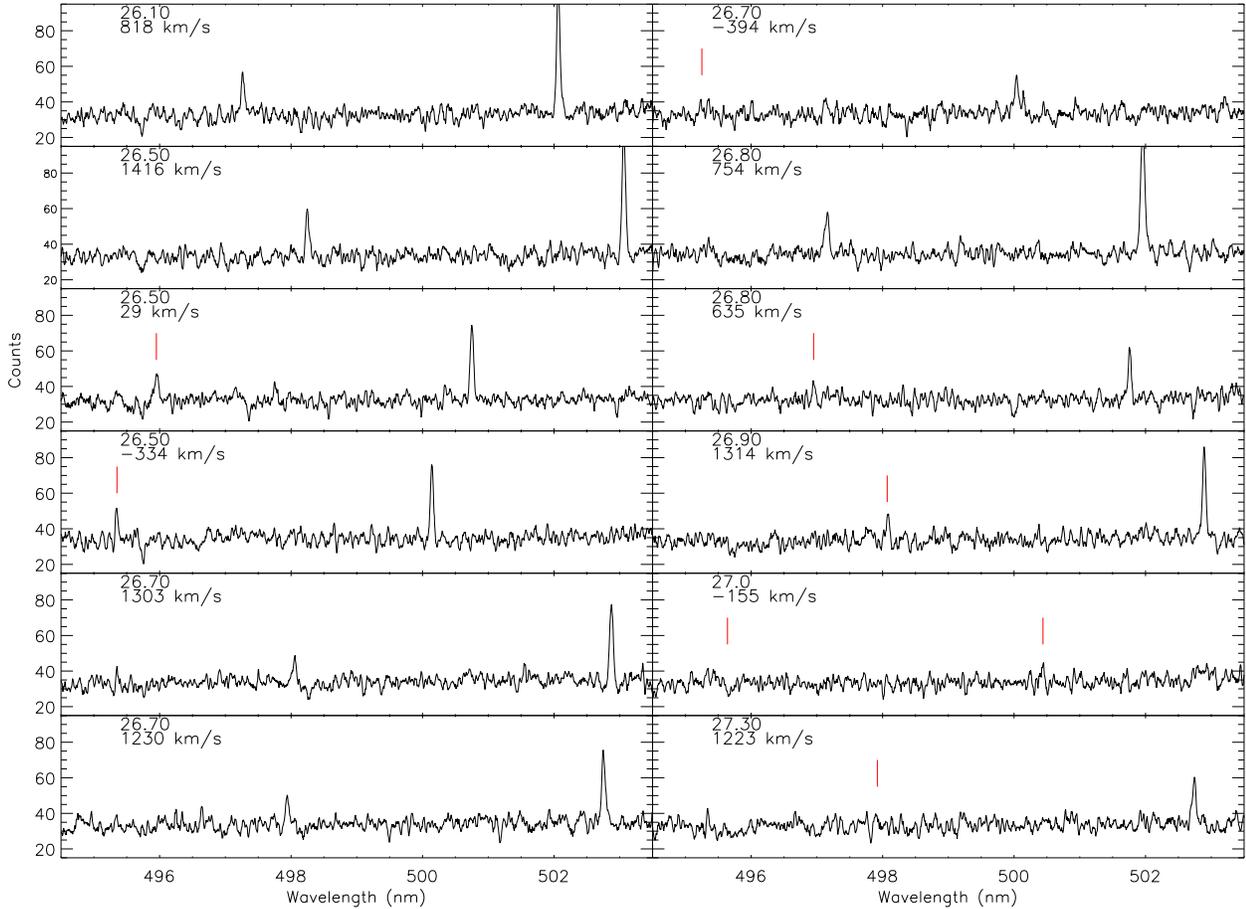}
\caption{Spectra for the confirmed PNs, ranked by magnitude m$_{5007}$
  and smoothed by a factor 7 to 0.035nm per pixel. m$_{5007}$ and the
  LOS velocities are labelled in the top left corner of each spectrum.
  The expected location of \Oiii$\lambda$4959 is shown by a red dash,
  in the cases where not immediately obvious.  The \Oiii 5007 line for
  the PNs with m$_{5007}$=27 and $\vel_{los}=-155\kms$ has a
  signal-to-noise ratio of 3 and this is our weakest believable
  detection. { A typical error in the velocities is 3 km/s; see
  Section~\ref{spectra}.}  }
\label{fig:spectra1}
\end{figure*}

A histogram showing the number of candidates versus number of
confirmed emission-line objects by magnitude is shown in
Figure~\ref{fig:completeness}. The photometric completeness limits
($\sim 90\%$) in the two photometric fields F4 ($m_{5007}$=26.6) and
F7 ($m_{5007}$=26.8) are shown as blue dotted lines. These photometric
completeness magnitude limits are defined as to where the
signal-to-noise over the entire photometric measurement is nine per
pixel or greater, corresponding to a photometric error of
approximately 0.12 magnitudes \citep{fcjd03}. The confirmation rate
for emission-line objects above the completeness limit is then 40-70\%
depending on the field. This is comparable with the results from A04
(30-80\% varying by field), and is a reasonable recovery rate given
the following effects.

Firstly, if the astrometry is not very precise or if some rotation
error is introduced in positioning the plate, then part or all of the
flux from some objects may miss the fibers. This is more serious for
faint objects as they will then not be detected above the
noise. Indeed, in Figure~\ref{fig:spectra1} the { total flux in the
\Oiii 5007 \AA\ line} is not clearly correlated with the magnitude
m$_{5007}$ of the source, indicating that fiber positioning might be
problematic. We measure the relative fluxes for the same \Oiii 5007
\AA\ detection in different frames and find that it can in fact vary
by up to a factor of two.

The likelihood of having false candidates above the completeness limit
is very low. Each candidate ICPN was hand-inspected, and the code that
finds the objects has been extensively tested on closer galaxies.  In
some cases (e.g., M51, \citet{dmf03}), there has been close to 100\%
recovery, using the same techniques as used to select the candidate
ICPNs here.

However, below the photometric completeness limit the uncertainties
are clearly much higher. Although many of the fainter objects are
still likely to be ICPNs and were hence targeted spectroscopically,
the probability for `spillover' (Aguerri et al. 2005\nocite{agm+05})
increases substantially. Due to the photometric errors in the [OIII]
fluxes some objects will be measured with a brighter flux than their
real flux.  If in addition their broad-band fluxes fall below the
limiting magnitude of the off-band image they will be selected as
ICPNs when they are in fact very faint continuum stars, due to the
fact that they will have erroneously large negative colours.

\subsection{Spectra of PNs and background emission line galaxies \label{spectra}}

Figure~\ref{fig:spectra1} shows all of the spectra for the confirmed
PNs, ranked by their photometric magnitude m$_{5007}$. For most of the
PNs brighter than m$_{5007}$=27 we also detect the second line
\Oiii$\lambda$4959. The expected location of \Oiii$\lambda$4959 is
shown by a red dash, where not immediately obvious by eye. The target
fields F7\_1 and F7\_2 overlap and have one source in common. The
independently measured velocities for this source agree to within
3\kms.

\begin{figure}
\includegraphics[scale=0.35,angle=90]{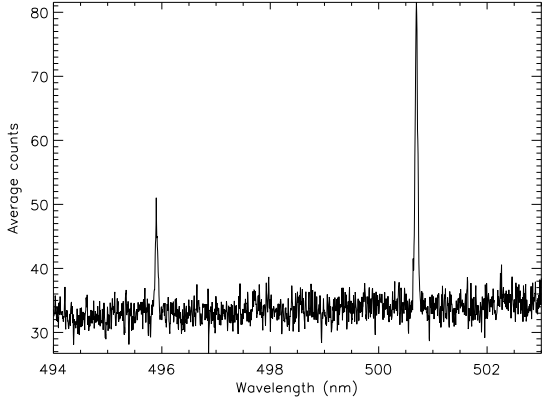}
\caption{Combined spectrum of all 12 identified PNs, Doppler corrected
to the rest-frame.}
\label{fig:stack}
\end{figure}

\begin{figure}
\includegraphics[angle=90,scale=0.6]{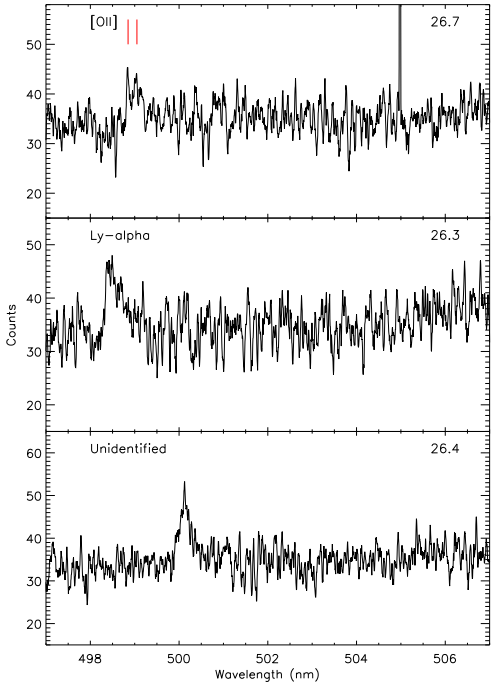}
\caption{Examples of the other emission-line objects present in the
sample, [OII], Ly$-\alpha$, and an unidentified broad emission line
which we speculate might be HeII at 1640\AA\ in a high-$z$ LBG, or
alternatively CIV\,1550 or [CIII]\,1909 from an AGN. The spectra have
been smoothed to 0.035 nm per pixel. For comparison to
Figure~\ref{fig:spectra1} the m$_{5007}$ magnitudes are shown in the
top right corner.}
\label{fig:spectra2}
\end{figure}

\begin{figure}
\includegraphics[scale=0.5]{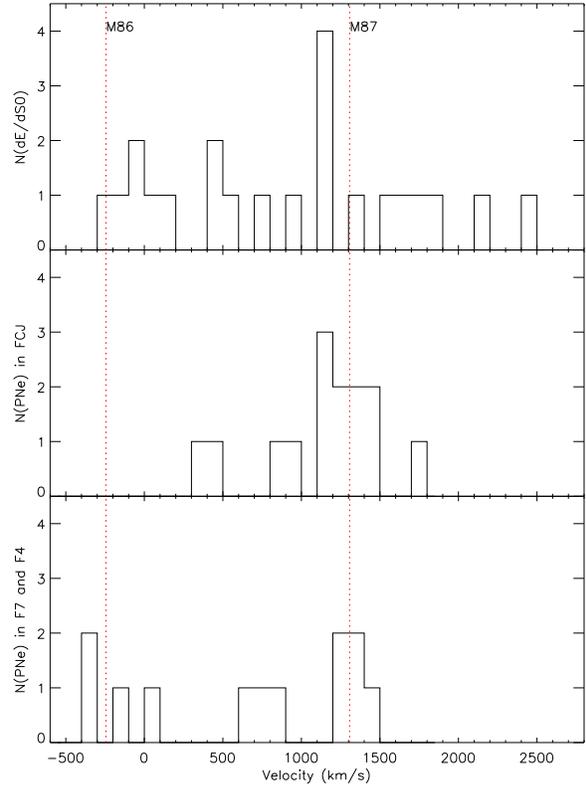}
\caption{Radial velocity histograms.  The bottom panel shows the
  velocity distribution of all identified PNs in the 3 new fields. The
  peak at $\sim1300\kms$ corresponds to PNs bound to the halo of
  M87. They have a mean velocity of $1297\pm35\kms$ and rms dispersion
  of $78\pm25\kms$. The middle panel shows the distribution of PN
  velocities in the previously surveyed FCJ field (A04), and the top
  panel shows the distribution of dwarf spheroidal LOS velocities in
  the same region of the Virgo cluster core, for comparison. The
  systemic velocities of M87 and M86 are shown with the dotted
  lines.}
\label{fig:vel_distrib}
\end{figure}

Our weakest believable detection has a total signal-to-noise ratio S/N=3. As
an additional check we create the average combined spectrum for the 12
identified PNs (Figure~\ref{fig:stack}) and measured the equivalent
width ratio of the two [OIII] lines. The ratio is 3 as expected if
all identifications are real.

In Figure~\ref{fig:spectra2}, examples of the other emission-line
objects present in the sample are shown: an \Oii\ doublet, an
asymmetric Ly$-\alpha$ line, and an unidentified broad emission line
which might be an AGN (for example CIV1550 or
CIII]1909). Alternatively, there is a possibility that the line is
HeII at 1640\AA\ in a LBG at higher redshift. Shapley et
al. (2003)\nocite{sspa03} discuss that this is sometimes seen as
nebular emission and also as much broader emission ($\sim1500\kms$)
possibly from stellar winds. The FWHM of the lines we see is about
$\sim200\kms$ (observed frame). The contamination rate of these other
emission-line objects is 7/19, or 37\%.

The PN emission lines are all resolved, and thus we have been able to
measure the expansion velocities of these PNs and to derive
information on the masses of the progenitor stars. This work is
presented in a companion paper \citep{adg+08}. Here we concentrate on
the kinematics, yielding information on the halo of M87 and the
assembly history of the Virgo cluster.

\section{LOSVD and projected phase-space\label{LOSVD}}

Figure~\ref{fig:vel_distrib} shows the distribution of velocities of
the newly identified PNs in the Virgo cluster core. 
The velocities have not been adjusted for a heliocentric correction, as
the observations were almost all taken close to the equinox and the
correction is within only $\pm3\kms$.

\begin{figure*}
\includegraphics[scale=0.8]{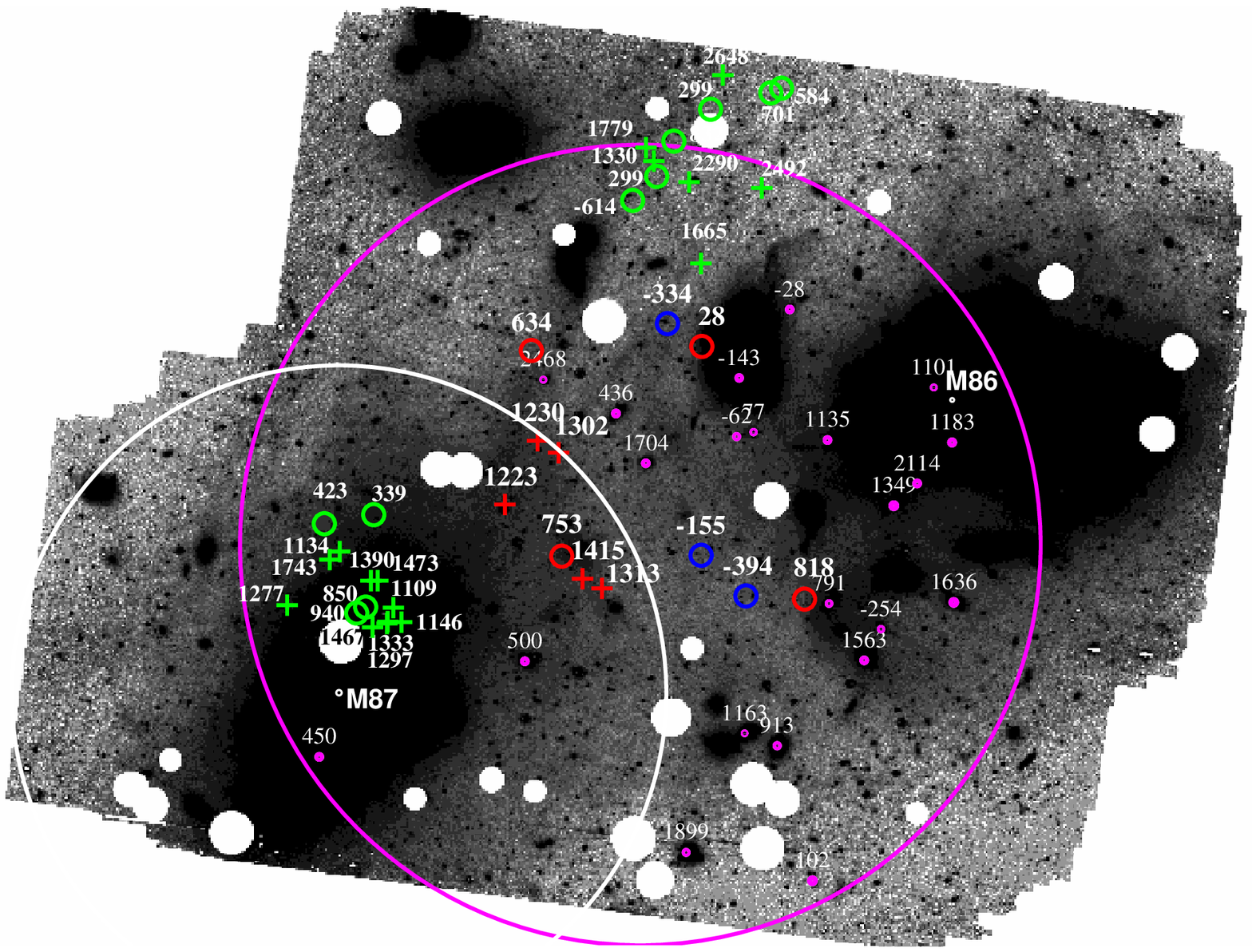}
\caption{Deep image of the Virgo cluster core showing the distribution
  of the intracluster light (\citet{mhfm05}). The spatial distribution
  of our spectroscopically confirmed PNs are overlaid. The A04 targets
  are shown in green. Our new targets are shown in red if redshifted
  with respect to Earth and blue if blueshifted. Objects with
  velocities higher than the mean velocity of Virgo (1100 \kms) are
  shown as crosses and those with lower velocities shown as circles.
  Dwarf spheroidals are marked as magenta dots. The velocities (in
  \kms) are labelled for all objects shown. The nominal `edge' of the
  M87 halo at $\sim \sim 160$ kpc is indicated with a white
  circle.  The pink circle has a 1.5 degree diameter and is centered
  on the projected midpoint of M87 and M86. North is up and East is to
  the left.}
\label{fig:vel_streams}
\end{figure*}

\begin{figure}
\includegraphics[scale=0.45]{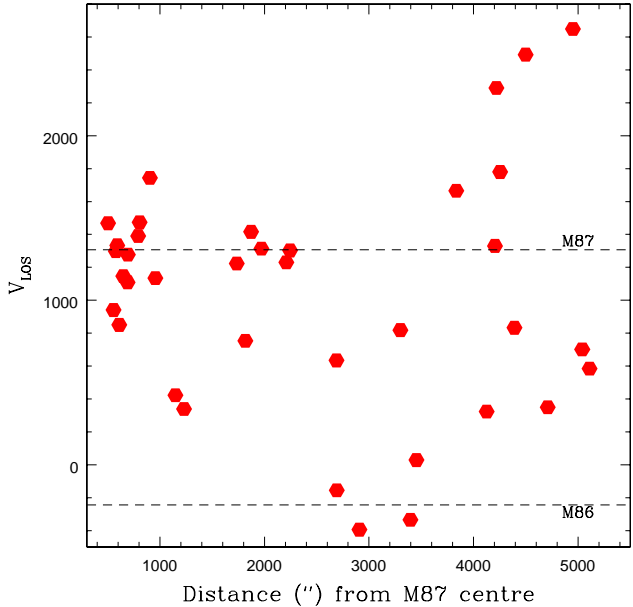}
\caption{ Distribution of line-of-sight velocity versus projected
  distance from the center of M87 for all spectroscopically confirmed
  PNs in the new fields as well as the FCJ and Core fields of A04.}
\label{fig:phasespace}
\end{figure}

{

For the subsequent analysis, we combine these velocities with the A04
sample in the FCJ and Core fields.  Figure~\ref{fig:vel_streams} shows
the location of these PNs on the deep image of the cluster core from
\citet{mhfm05}, and Figure~\ref{fig:phasespace} shows their
distribution in the projected phase-space plane defined by projected
distance from M87 center and line-of-sight velocity.

In the phase-space diagram Figure~\ref{fig:phasespace}, we can
identify two regions with very different characteristics: For
projected distances $R<2400"$ most of the PNs are strongly clustered
around the systemic velocity of M87, $\vel_{sys}=1307$ \kms. By
contrast, for $R>2400"$, the PN velocities spread widely over a
velocity range more typical for the Virgo cluster. From the latter,
intracluster region we see a string of low PN velocities ($800$-$400$
\kms), extending inwards to the upper FCJ field (see
Fig.~\ref{fig:vel_streams}).

In the FCJ field at projected distance $R<1300"$ there are two of
these intracluster outliers at $\sim 400$ \kms. The remaining PNs are
distributed symmetrically around $\vel_{sys}$ and have mean velocity
$1276\pm71\kms$ and velocity dispersion $\sigma = 247$ \kms (A04);
their velocity distribution is shown in the middle panel of
Fig.~\ref{fig:vel_distrib}.

In the combined new F7/F4 fields at $R\sim 2000"$ we find five PNs
tightly clustered around $\vel_{sys}=1307$ \kms; these have mean
velocity $1297\pm35\kms$ and an rms dispersion of $78\pm25\kms$.  At
comparable radii there are two additional PNs with velocities of 753 and
634 \kms; compared to the previous five, these are $7 \sigma$ and $8
\sigma$ outliers. It is unlikely that one or two of these outliers are
part of the same (very asymmetric) distribution as the five PNs
clustered around $\vel_{sys}$. By contrast, they fit naturally into
the stream leading from the FCJ outliers all the way into the ICL.  We
therefore identify as PNs bound to the M87 halo only those PNs which
are clustered around the systemic velocity of M87.  These are confined
to radii $R<2400"$.

The M87-bound PNs in the FCJ and combined F7/F4 fields are located at
mean projected radii of 60 and 144 kpc, respectively. They correspond
to the narrow peaks in the line-of-sight velocity distributions
(LOSVD) in the lower and middle histograms in
Fig.~\ref{fig:vel_distrib}.

Outside $R=2400"$ in Fig.~\ref{fig:phasespace} we find PNs at larger
relative velocities to M87, with an approximately uniform distribution
in the range $-300$ to $2600$ \kms. Those in the radial range
$2400"<R<3600"$ (the F7/F4 field) are confined to negative velocities
with respect to M87.  These are probably encroaching stars from M86
and other Virgo components\footnote{They cannot be in the Local Group:
  the faintest PNs in the SMC have m$_{5007}$=23 \citep{jacoby02} so
  are still much brighter than the brightest Virgo PNs at
  m$_{5007}$=26.3.}. By contrast, the PNs further than $3600"$ from
M87 (in the Core field) show a broad distribution of velocities, more
characteristic of the cluster as a whole (see A04).

In the middle and bottom panels of Fig.~\ref{fig:vel_distrib}, the ICL
PNs show up as approximately flat velocity distributions in their
velocity range, besides the peak of velocities from PNs bound to M87.
A flat distribution of velocities in addition to the peak near M87's
systemic velocity is also seen in the LOSVD of the dwarf spheroidal
galaxies in the same region of the Virgo cluster core \citep{bpt93}
which is shown in the top panel of Figure~\ref{fig:vel_distrib}.
However, for the dwarf galaxies the flat velocity distribution extends
to significantly more redshifted velocities, indicating that the dwarfs
and ICL PNs kinematics can only partially be related. 

}

\section{The M87 Halo\label{M87}}

{

We have seen from the phase-space diagram in Fig.~\ref{fig:phasespace}
that the PNs in the FCJ and F7/F4 fields divide into two components, one
associated with the halo of M87, and the other with the unbound Virgo
ICL.  All PNs found around the M87 systemic velocity are within
$R=161$ kpc projected radius from the galaxy's center.  In the following
subsections we combine our velocity measurements with the kinematic
data in the literature, and discuss the rotation, velocity dispersion
profile, dynamics and truncation of the outer M87 halo. Finally we
consider possible origins of the truncation.
}

\subsection{Rotation of outer halo?} 

First we ask whether there is any evidence in our data for rotation in
the outer halo of M87. For the globular cluster (GC) system of M87,
\citet{cr97} inferred a rotation of about 100~\kms\ for $R\lta 35$
kpc, approximately about the minor axis of the galaxy intensity
isophotes, using spectra of low resolution with errors for the GC
velocities of order 100~\kms.  Cohen (2000) added new data for GCs in
the halo at $24<R<43$ kpc with smaller errors (typically
$\sim50\kms$), and inferred a rotation of 300\kms.  \citet{cmh+01}
carried out an independent analysis using a new spectroscopic and
photometric database \citep{hcb+01} partly based on that of
\citet{cr97} and \citet{coh00}, and similarly find
$\sim160\kms$. \citet{cmh+01} found that the metal rich GCs rotate
everywhere about the photometric minor axis of the galaxy, while the
metal poor GCs have a more complex behaviour: they rotate about the
photometric minor axis of the galaxy between $15<R<40$ kpc, and about
the major axis at $R<15$ kpc.

If the PN population in the outer halo of M87 also rotated about the
galaxy's photometric minor axis, similarly to the M87 GC system at
$15<R<40$ kpc, we should see a signature along the major axis of the
galaxy, that is, the mean velocities of the PN LOSVD peaks associated
with M87 in our two pointings FCJ \citep[A04; this is F3 in][]{fcjd03}
and F7/F4 (Figure~\ref{fig:ourfields}) should be offset from the
systemic velocity of the galaxy (1307 \kms). From the extrapolated fit
of \citet{cmh+01} to the GC radial velocities we would expect a
negative constant offset of about $160$ \kms at both field positions,
i.e., a mean velocity of $\sim 1150$ \kms.

{ 

  For the M87 sample of PNs identified in the phase-space diagram in
  Fig.~\ref{fig:phasespace}, we find a mean velocity of
  $\bar{\vel}_{\rm rad}=1297\pm35\kms$ in the new field (F7/F4) at
  mean projected radius $R=144$ kpc.  In the previously studied field
  (FCJ; A04) the mean velocity is $1276\pm71\kms$ at mean $R=60$ kpc.
  Thus we see no evidence for rotation of the outer stellar halo
  around the galaxy's minor axis in either the PN sample at $R=60$ kpc
  or at $R=144$ kpc.  The rotation seen in the GCs may thus suggest
  that they do not trace the main stellar population of M87, or that
  they are contaminated with IC GCs with a LOSVD similar to the
  encroaching stars of M86 as seen in Fig.~\ref{fig:phasespace}
  \citep[see also Fig.~1 of][]{cmh+01}.  We have not surveyed fields
  along the minor axis, so we cannot check with PNs whether there is
  rotation about the major axis.

}

\subsection{Velocity dispersion profile}

\begin{figure}
\includegraphics[scale=0.4]{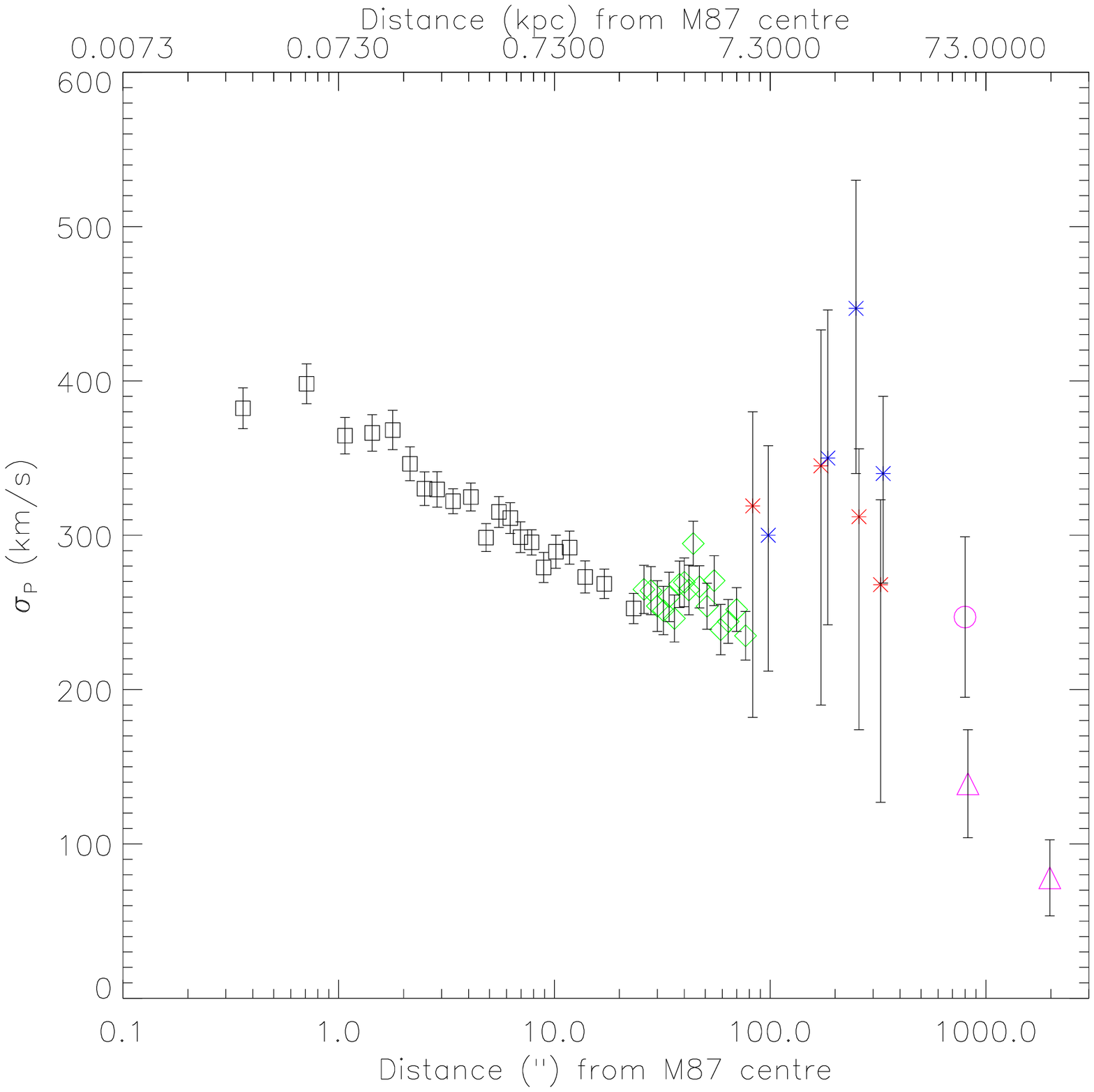}
\caption{Velocity dispersion profile of M87, including stellar
  velocity dispersions from absorption-line spectra and discrete LOS
  velocity measurements from globular cluster and PN data. The squares
  are data points from \citet{vdM04}, the green diamonds are based on
  \citet{st96}, and the red and blue stars are velocity dispersions
  for the metal-rich and metal-poor GC samples of \citet{cmh+01},
  respectively. The magenta circle is the PN velocity dispersion
  determined in A04 and the magenta triangles are the PN dispersion
  values from this paper. These last two points are approximately
  along the major axis of the outer isophotes, which have ellipticity
  $\epsilon\simeq0.43$.}
\label{fig:veldispprof}
\end{figure}

With the new data, we can also now plot the velocity dispersion
profile of M87 all the way out to $R=144$ kpc from the centre of the
galaxy.  Figure~\ref{fig:veldispprof} shows this as a function of
projected radius from M87 centre. In the inner regions ($R<25\arcsec$)
we use the G-band absorption line measurements from the integrated
stellar light of van der Marel (1994)\nocite{vdM04}. In the region
$25\arcsec<R<80\arcsec$ we use stellar velocity dispersions from
Sembach \& Tonry (1996). As these authors discuss, their data is
systematically offset from most other datasets by $7-10\%$, due to
using a larger slitwidth. \citet{rk01} calculate that this amounts to
an additional instrumental dispersion of 183\kms\ and so we adjust the
velocity dispersion by this amount (subtracting in quadrature) to
bring it in line with the van der Marel dataset. We take the average
of the velocity dispersions at each positive and negative $R$, assuming
symmetry with respect to the galaxy's center.  We exclude the Sembach
\& Tonry data in radial bins beyond 80\arcsec\ as there is a
discrepancy between the velocity dispersions at the corresponding
opposite positions in radius along the axis and furthermore the values
in those bins have large error bars. This may be due to low S/N in the
outer part of the galaxy where the surface brightness is low and/or
real anisotropies in the velocity dispersions. Either way we judge it
better to exclude these data points as they are less trustworthy.

\begin{figure}
\includegraphics[scale=0.4]{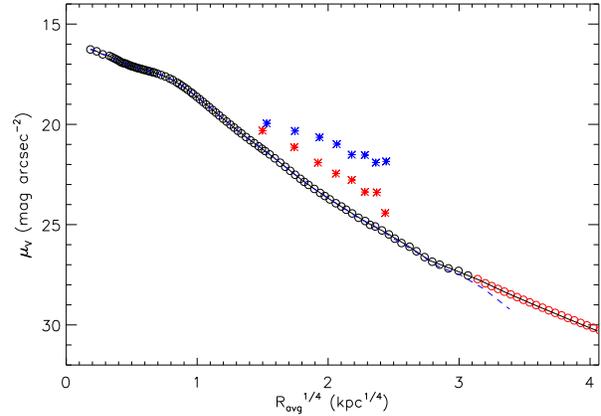}
\caption{V-band surface brightness profile for M87 from
  \citet{kormendy+08} shown with black circles along average ellipse
  radii $R_{avg}$ of the isophotes. Red circles show the extrapolated
  S\'{e}rsic fit. The full line shows the reprojected
  three-dimensional luminosity model obtained from the extrapolated
  data, and the dashed line shows the reprojection of the luminosity
  model when truncated { at average ellipse radius} $R_{trunc} =
  149$ kpc. For comparison, the number density profiles of the red and
  blue globular cluster populations from \citet{cmh+01} are also shown
  { with arbitrary scaling} as red and blue stars, respectively. }
\label{fig:M87_sb}
\end{figure}

We also show in Figure~\ref{fig:veldispprof} the data of
\citet{cmh+01} for the metal-rich GCs out to 380\arcsec. We exclude
the outer bins (380\arcsec -- 635\arcsec) where the error bars are
close to 100\% and therefore do not constrain the shape of the
velocity dispersion profile in any way. The metal-poor GC system is
more spatially extended than its metal-rich counterpart \citep[see
Figure~\ref{fig:M87_sb} below and][]{cmh+01} and may be composed of
accreted and/or infalling remains of `failed' or disrupted dwarfs:
their velocity dispersion data are also shown in
Figure~\ref{fig:veldispprof}, but do not trace the velocity
dispersions of the M87 stars. We also note from Figure 1 of
\citet{cmh+01} that the GC sample is likely to contain intracluster
GCs just as our PN sample contains ICPNs, requiring a careful analysis
of the GC LOSVDs.

Finally, the two outermost velocity dispersion points are from
planetary nebulas presented in A04 and this paper. We note that when
we bin the A04 data to be consistent with the binning of the velocity
distribution in this work (100\kms\ bins) the peak around M87 is
resolved into a somewhat narrower peak of 9 objects, with two lower
velocity and one higher velocity outliers
(Figure~\ref{fig:veldispprof}). { The mean and rms velocity of this peak
of $\bar{\vel} = 1264$ \kms\ and $\sigma = 247$ \kms from A04 then
change to $1292\pm46\kms$ and $139\pm23\kms$ respectively.  It is
possible that the larger value of $\sigma_{PN}$ obtained by A04 could
be due to the inclusion of some ICPNs from the component with uniform
LOSVD seen in Figure~\ref{fig:vel_distrib}. We carried out $\chi^2$
tests\footnote{We carried out a $\chi^2$ test for the (FCJ;A04) sample
  and i) a broad Gaussian ($\bar{\vel} = 1264$ \kms\ and $\sigma =
  247$ \kms), ii) a uniform distribution plus a narrow Gaussian
  ($\bar{\vel} = 1264$ \kms\ and $\sigma = 139$ \kms). Because of the
  limited statistics of the PN sample in this field, the results
  depend on the velocity range chosen for the test. In a $700 -
  1650$ \kms\ range, the two distributions fit the data equally well
  with $80\%$ probability, while the broad Gaussian is ruled out in
  a $350 - 1650$ \kms\ range.}  but could not distinguish between
both interpretations.  The dynamical model discussed below favors the
high value of $\sigma_{PN}$ at $R=60$ kpc.

In any case, the PN velocity measurements show that the halo of M87
becomes colder at larger radii: the velocity dispersion decreases to
78\kms\ at $R \sim 140$ kpc.

}

\subsection{Truncation of the M87 stellar halo\label{truncation}}

In the FCJ field, there are M87 halo PNs detected throughout, but in
the F7/F4 fields, the PNs around the systemic velocity of M87 (=1307
\kms) appear to be found only within a projected radius of $R= 161$
kpc (see the spatial distribution of the spectroscopically confirmed
PNs in Figure~\ref{fig:vel_streams}). At projected radii $R > 161$ kpc
from the center of M87, we find only the encroaching stars of M86 and
other Virgo components. We now investigate whether this spatial and
velocity segregation is significant and indicates that the M87 stellar
halo is truncated.
 
\begin{figure}
\includegraphics[scale=0.30]{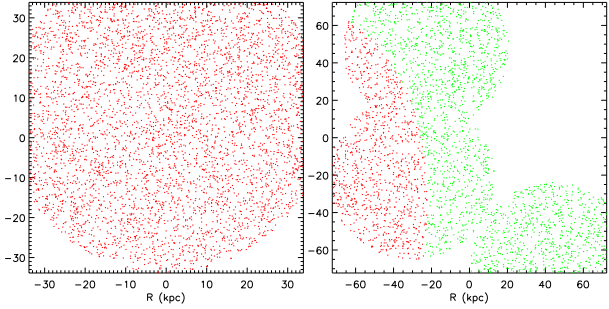} 
\caption{Left: The red region is the intersection between the photometric
  field FCJ and the FLAMES (FCJ) pointing.
  Right: The red region is the part of the intersection between the
  photometric field F7 from Feldmeier et al. (2003) and the regions
  jointly covered by the F7\_1, F7\_2 and F4 FLAMES circular
  pointings, which is within the isophote with an average ellipse
  radius, $R_{trunc}=149$ kpc. This covers the region containing PNs
  with velocities bound to M87. The green region is the part of the
  intersection between the F7 region and the FLAMES pointings as
  before, but outside $R_{trunc}$, where no PNs associated with the
  M87 velocity peak are found.}
\label{area:red:green}
\end{figure}

\citet{kormendy+08} present a composite V-band surface-brightness
profile for M87 out to 135 kpc along the semi-major-axis. This is
shown with black circles in Figure~\ref{fig:M87_sb}, with the
semi-major axis of each isophote replaced by the average ellipse
radius, $R_{avg} = (ab)^{1/2}$ of the isophote. We will use the latter
in the construction of the spherical Jeans models in
\S~\ref{sec:dynamics}.  \citet{kormendy+08} fit a S\'{e}rsic profile
to the semi-major axis profile excluding the central core and the last
two data points (which may have a significant ICL contamination), and
obtain the following best-fit parameters: $\mu_e = 25.71$, $R_e = 704"
= 51.2$ kpc, $n = 11.885$. We use this S\'{e}rsic fit to the surface
brightness profile (see Fig.~\ref{fig:M87_sb}) to compute the
luminosity of the M87 halo at radii outside the available photometry.
We note that the description of M87 as a classical E0 or E1 galaxy is
based on short exposure optical images, while in deep images its
isophotes show marked eccentricity. For the extrapolation we assume an
ellipticity $\epsilon=0.43$ and position angle (PA$=-25^\circ$), based
on the outer parts of the ellipticity and position angle profiles in
\citet{kormendy+08}. We then evaluate the M87 halo luminosities in the
regions of overlap between the photometric and spectroscopic fields,
in which, respectively, the photometric identification and
spectroscopic follow-up of the PNs was carried out. These are shown by
the colored regions in Figures~\ref{area:red:green}. For ellipticity
$\epsilon=0.43$, the PNs belonging to M87 appear to be found within an
average ellipse radius $R_{avg}=149$ kpc. We use the isophote
corresponding to this radius to demarcate the region containing PNs
belonging to M87 (in red in Fig.~\ref{area:red:green}) from the region
containing no PNs belonging to M87 (in green in
Fig.~\ref{area:red:green}). Table~\ref{tab:PNLTOT1} gives the areas
and the V-band luminosities of the various regions obtained. On the
basis of the S\'{e}rsic fit, the ratio of the M87 luminosity in the
F7-green and in the F7-red area is $0.92$.

From the number of spectroscopically confirmed M87 PNs in the F7-red
area (i.e., 5), and the ratio of the M87 halo luminosities from the
S\'{e}rsic fit (i.e., 0.92), we can then predict the expected number
of PNs at the M87 systemic velocity in the F7-green area, to be $ 5
\times 0.92 = 4.6$ PNs.  The observational result of zero M87 PNs
detected in the F7-green area thus implies a truncation of the M87
stellar halo at $R_{avg} > 149$ kpc, at a $\simeq 2\sigma$ level.
This radius tells us the location of the outermost PN in terms of the
average ellipse radii and we now refer to it as the truncation radius
$R_{trunc}$.

\begin{table}[h]
\caption{M87 halo PNs and sampled luminosities.}
\begin{tabular}{lccc}
\hline 
Field    &  N$_{spectr}$  &  Area        & L             \\
         &                &  kpc$^2$     & L$_{V,\odot}$ \\
\\
FCJ red  &   12           &    4157      &  $8.7\times 10^9$   \\
F7  red  &   5            &    4349      &  $1.3\times 10^9$   \\
F7 green &   0            &    8909      &  $1.2\times 10^9$   \\
\hline\\
 
\end{tabular} \\
\label{tab:PNLTOT1}
\normalsize
\end{table}

This is a surprising result: the numerical simulations of
\citet{ans06} find that the luminous halos around isolated galaxies
should extend to the virial radius, i.e., to several hundred kpc and
well beyond their traditional luminous radius. It is therefore not
obvious why there should be a truncation of the M87 stellar halo, see
Section~\ref{origint}.  One question is whether the truncation we see
occurs only at the targeted $P.A.$ within an opening angle $\delta
P.A.$, say, rather than at all azimuths, and whether when azimuthally
averaged, the halo light distribution would extend to larger radii.
This could be the case if we had reached the radii where the stellar
halo of M87 has a significant amount of substructure, similar to the
outer Milky Way halo \citep{bell+08}. For example, we might explain
the spatial segregation of the M87 halo PNs in terms of a cold stellar
shell at our field position, followed by a steeper surface brightness
profile at those $P.A.$ whereas at other $P.A.$ the profile would be
shallower. This would also explain the small $\sigma_{PN}$ that we
measure in our fields at $R=144$ kpc, as the stars populating shells
are near to the apocenters of their elongated orbits.

To assess this we must reconsider the photometric structure of M87.
Within a semi-major axis of $\sim 160$ kpc and $\mu_V \lta 27.5$, the
surface brightness distribution around M87 is well approximated by an
extended envelope with $c/a \simeq 0.57$ \citep{mhfm05, kormendy+08}.
This ellipsoidal component includes the diffuse fan of stellar
material, which extends along the projected southeast semi-major axis
out to $\sim 100$ kpc \citep{arp71,wbm97}, but is otherwise fairly
smooth.  At larger radii and fainter surface brightnesses, the light
distribution is a superposition of the outer halo of M87 and the ICL
and is brightest in the range of $P.A.$ towards M86 where our target
fields are.  At these radii it { does show} irregular features and
some radial streamers, and our fields are large enough to include
several of these. In fact, some of our outermost M87 PNs may be close
to an arc-like feature in the \citet{mhfm05} image beyond which little
light is seen. However, both the earlier results of \citet{wbm97} who
reported the apparent lack of sharp-edged fine structures around M87,
and our independent inspection of the \citet{mhfm05} image near M87,
provide no evidence for a large number of ``shell-like'' features at
various azimuths and radii around M87. This is true both inside and
outside our truncation isophote, and in particular around $R \sim
60$ kpc, where the PN data already indicate a falling velocity
dispersion profile (see Fig.~\ref{fig:veldispprof}).

Further investigation of the extended luminosity distribution around
M87 would require quantitative photometry of the deep image of
\citet{mhfm05}, and a large-area and wide-angle PN velocity survey to
separate the outer halo of the galaxy from the ICL with better
statistics.

In what follows, we follow an { independent} approach and test the
hypothesis of a truncated stellar halo in M87 dynamically. We will
verify whether we can make a dynamical model for M87 in which the
stellar velocity dispersion reaches low values everywhere around M87,
and the total gravitational potential is traced by the X-ray emission of
the hot gas.

\subsection{The mass distribution and anisotropy in the M87 outer halo
\label{sec:dynamics}}

{ The smooth photometric and X-ray emission profiles indicate that
the outer halo of M87 is in approximate dynamical equilibrium.}
With the extended velocity dispersion profile we are now able to
create dynamical models of M87 to infer the orbital structure in the
outermost halo. In a spherical system, the intrinsic velocity
dispersions of a population of stars with density $j$ moving in a
potential $\Phi$ are related by the second-order Jeans equation
\begin{equation}\label{eq:jeans}
\frac{d}{dr}\left(j(r)\sigma_r^2(r)\right) 
    + \frac{2\beta}{r}j \sigma_r^2(r)
    + j(r)\frac{d\Phi}{dr} = 0
\end{equation}
where the anisotropy parameter $\beta(r) = 1 -
(\sigma_{\theta}(r)/\sigma_r(r))^2$ quantifies the orbital structure
of the system. Its value spans a range between $- \infty$ signifying
purely tangential orbits and $1$ signifying purely radial
orbits. These intrinsic quantities are now in terms of the
three-dimensional radius, $r$.

Therefore if we know the potential $\Phi$, density of stars $j$ and
assume a radial dependence for the anisotropy, we can solve for the
intrinsic radial ($\sigma_r$) and tangential ($\sigma_{\theta}$)
velocity dispersion profiles. These can then be projected and compared
with the observed projected velocity dispersion profile to fix the
parameters of the assumed anisotropy profile. Under the spherical
assumption, the radii of the observables will be the average ellipse
radii $R_{avg}$ introduced in \S~\ref{truncation}.

One method of deriving the potential of a galaxy is to use electron
density and temperature profiles obtained from X-ray measurements of
the thermal bremsstrahlung emission from the hot interstellar medium
surrounding massive elliptical galaxies. If the gas is relatively
undisturbed then we can assume that the gas is in hydrostatic
equilibrium and thus derive the potential.

\begin{figure}
\includegraphics[scale=0.4]{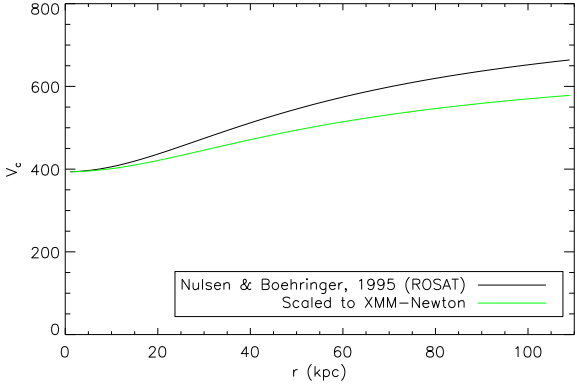} 
\caption{Circular velocity profiles for M87 from X-ray data. }
\label{fig:M87_vcirc}
\end{figure}

\citet{nb95} use $ROSAT$ data and a maximum likelihood
method to deduce the most likely mass profile in the Virgo cluster
core, extending from the centre of M87 out to 300 kpc. They
parametrize this profile with a model composed of two (approximate)
isothermal mass distributions, one attributed to the mass of M87 with
a mass per unit length $\mu = 3.6\times 10^{10} M_{\odot}$ kpc$^{-1}$
and the other to the dark matter of the cluster with a mass per unit
length $M_0/a = 12.4\times 10^{10} M_{\odot}$ kpc$^{-1}$ and a core
radius $a = 42$ kpc. This parametrization is given as
\begin{equation}\label{eq:mass}
M(r) = \mu r + M_0 [(r/a) - \arctan(r/a)].
\end{equation}
This mass profile is related to the potential through $ r d\Phi(r) /dr
= \vel_c(r)^2/r = GM(r)/r^2$, where $\vel_c(r)$ is the circular velocity at
$r$. It is more easily calculated than the potential but also
independent of distance. The circular velocity profile is shown in
Figure~\ref{fig:M87_vcirc} and increases from a minimum circular
velocity of 393 kms$^{-1}$ to eventually a maximum velocity of 830
kms$^{-1}$. For comparison the same profile is shown with $M_0$
adjusted so that the maximum velocity reached is 700 kms$^{-1}$, the value
suggested by the extrapolation of newer observations taken with {\it
XMM-Newton} \citep{mab02}.

The density of the stars was obtained through the deprojection of the
surface-brightness profile of M87. To obtain the intrinsic luminosity
density, we adopt the \citet{kormendy+08} profile (also missing the
last two data points), and using their S\'{e}rsic fit, extrapolate to
very large radii, see \S~\ref{truncation}; then we employ the standard
deprojection formula. In Figure~\ref{fig:M87_sb}, the S\'{e}rsic
extrapolation is shown by the red circles and the reprojection of the
intrinsic luminosity density is shown by the solid black line which
follows the circles very well.

\begin{figure}
\includegraphics[scale=0.4]{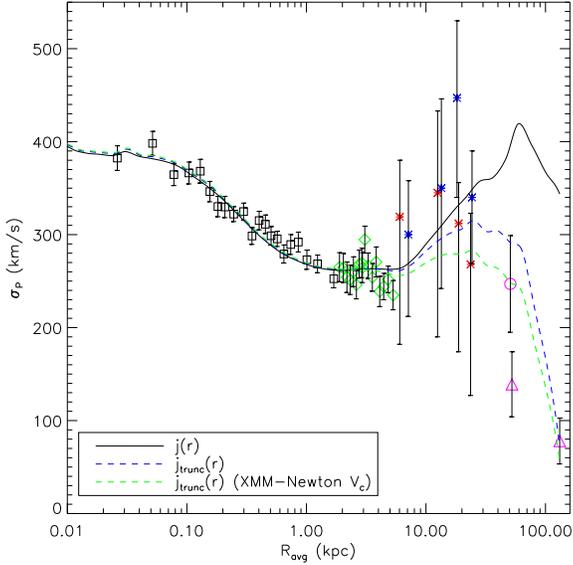} 
\caption{Velocity dispersion profiles derived for Jeans models with
  spherical symmetry and surface brightness profiles as in
  Figure~\ref{fig:M87_sb}; see text. The velocity dispersion data
  points are shown at their average ellipse radii $R_{avg}$, computed
  with the ellipticity profile.  For the globular cluster velocity
  dispersions this is not possible; these are not used in constructing
  the Jeans models.  Symbols are as in Fig.~\ref{fig:veldispprof}.}
\label{fig:M87_vdp}
\end{figure}

Finally, the radial dependence we adopt for the anisotropy profile is
given by
\begin{equation}\label{eq:beta}
\beta(r) = \beta_m \frac{r + r_1}{r + r_2}
\end{equation}
where $r_1<r_2$, $\beta_m$ is the maximum anisotropy reached at large
$r$, $\beta_m r_1/r_2$ is the
minimum anisotropy and $r_2$ represents the scale of the transition
from the minimum anisotropy to the maximum anisotropy. This solves the
Jeans equation (\ref{eq:jeans}) as a first-order differential equation
with the integrating factor
\begin{equation}\label{eq:if}
\textrm{I.F.} = j(r) r^{2\beta_m x} (r + r_2)^{2\beta_m(1-x)}
\end{equation}
{ where $x = r_1/r_2$ so that 
\begin{equation}\label{eq:rsig}
  \sigma_r^2(r) = \frac{1}{\textrm{I.F.}} \int_r^{\infty} \textrm{I.F.} 
      \frac{\vel_c^2}{r'} dr'.
\end{equation}
Then the projected velocity dispersion, $\sigma_P$, is obtained by
projecting the intrinsic velocity dispersions along the line of sight.
These integrals were determined numerically.

To obtain a dynamical model for the S\'ersic light distribution in the
potential implied by Fig.~\ref{fig:M87_vcirc}, we fixed the minimum
and maximum anisotropy using constant anisotropy models and then
employed a $\chi^2$ minimization approach to deduce the best-fit $r_2$
for the solution in eq.~\ref{eq:rsig}.  This minimization takes into
account the long-slit data, the $\sigma_{PN}$ point at 60 kpc from
A04, and the new $\sigma_{PN}$ point at 144 kpc, but not the globular
cluster data. The best-fit model is shown by the solid black line in
Figure~\ref{fig:M87_vdp}. It fits the data very well within 6 kpc but
it is unable to reproduce the low PN velocity dispersions in the outer
parts, at $R_{avg}=52$ kpc and $R_{avg}=131$ kpc, where it would predict
LOSVDs with dispersions of 350-400 \kms.}

On the basis of the results in \S~\ref{truncation}, we assume now that
the galaxy's intrinsic luminosity density is truncated at
$r=R_{trunc}=149$ kpc (i.e., in a spherical system the intrinsic
truncation radius is the same as the projected truncation radius).
The reprojection of this truncated intrinsic luminosity density is
also shown in Figure~\ref{fig:M87_sb} by the blue dashed line.  We
construct a Jeans model for the truncated luminosity distribution in
the same way as above.  Finally, to check the influence of the assumed
potential on the results, we have evaluated one further model, also
assuming a truncation in the luminosity density but using the circular
velocity that was adjusted to the analysis of {\it XMM-Newton}
observations in \citet{mab02}.  The corresponding Jeans models are
shown in Figure~\ref{fig:M87_vdp} with the blue and green dashed lines
respectively, with the second model dipping slightly lower in the
outer parts, reflecting the lower potential in this region.

\begin{figure}
\includegraphics[scale=0.4]{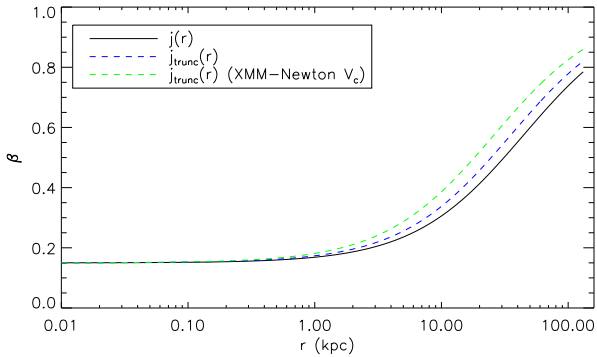} 
\caption{ Anisotropy profiles for the best-fit models: they imply
a mildy radially anisotropic orbital distribution ($\beta \approx
0.15$) in the centre that becomes highly radially anisotropic in the
outer halo ($\beta \approx 0.8$).}
\label{fig:M87_bp}
\end{figure}

Both truncated models behave as the untruncated S\'{e}rsic model in the
center, but fall much more steeply at radii $R_{avg} > 30$ kpc, thus
being able to reproduce the outermost $\sigma_{PN}$ data points at
$R_{avg}=52$ and $R_{avg}=131$ kpc (corresponding to projected radii
$R=60$ and $R=144$ kpc).  At $R_{avg}=52$ kpc, the truncated models
predict a velocity dispersion which favours the higher value of
$\sigma_{PN}$, i.e., 247 km~s$^{-1}$. Figure~\ref{fig:M87_bp} shows
that the best-fit models imply a mildy radially anisotropic orbital
distribution ($\beta \approx 0.15$) in the centre that becomes highly
radially anisotropic in the outer halo ($\beta \approx 0.8$).

We conclude that, under the assumption of spherical symmetry, the
Jeans models can only reproduce the low PN velocity dispersion
measurements in the outer halo of M87 at $R_{avg} > 30$ kpc only with
a truncation of the intrinsic luminous density.

{ In principle, this dynamical argument could be circumvented if at
  the position of our outer fields the stellar halo of M87 was
  strongly flattened along the line-of-sight. In this case
  $\sigma_{PN}$ could be low at these radii independent of a
  truncation.  However, at $\sim 150$ kpc radius such a flattening is
  likely to be local and would have arisen from the geometry of
  accretion, rather than signifying an angular momentum supported
  global structure collapsed from even $\lambda^{-1}\simeq 20$ times
  further out. Thus the well-mixed, three-dimensional stellar halo of
  M87 would then have ended at even smaller radii. Also note that in
  this case we could still not explain the lack of PNs at the M87
  systemic velocities for radii greater than $R_{trunc}$ (see
  Fig.~\ref{fig:phasespace} and the discussion in
  \S~\ref{truncation}.)}

\subsection{On the possible origin of a truncated stellar halo in M87
  \label{origint}}

{ Summarizing the last two sections, there are two independent and
mutually consistent pieces of evidence that the stellar halo of M87
ends at $R_{trunc}\simeq 150$ kpc: the lack of PNs around the systemic
velocity of M87 beyond this radius, and the very low velocity
dispersion in the outer halo.

In well-mixed, dense galaxy clusters it is expected that galaxies are
tidally truncated by the cluster's tidal field
\citep{merritt84,ghigna+98}. The tidal effects are strongest near the
cluster core radius and the approximation $r_{\rm tidal}\simeq r_{\rm
  peri} \sigma_{\rm halo} / \sigma_{\rm clus}$ for the tidal radius is
found to work well.  Here $r_{\rm peri}$ is the pericenter of the
galaxy's orbit in the cluster, and $\sigma_{\rm halo}$ and
$\sigma_{\rm clus}$ are the velocity dispersions of the galaxy halo
and the cluster, respectively. The tidal truncation of the dark matter
halos of galaxies has been detected with combined strong and weak
lensing observations in several dense clusters
\citep{natarajan+98,natarajan+02, limousin+07,halkola07}. Tidal radii
of between $15$ and $60$ kpc have been found, in agreement with
predictions.

The case of M87 is not so simple though. M87 is at the center of at
least a subcluster potential well, traced by the X-ray emission and
the dark matter mass profile derived from it \citep[][see
Fig.~\ref{fig:M87_vcirc}]{nb95,schindler+99}.  A galaxy at the center
of its cluster experiences a symmetric gravitational field from the
cluster dark matter, and is consequently not gravitationally truncated
\citep{merritt84}.  On the other hand, M87 has a relative motion of
$\sim 300$ km/s with respect to the galaxies in the cluster core
\citep{bpt93}, and the galaxy distribution in the core is complicated
and not centered on M87, containing a strong concentration around
M84/M86 \citep{bts87}.  It is possible that the M84/M86 concentration
including the associated dark matter exerts a significant tidal field
on M87.  There is no obvious feature in the density of the X-ray
emitting gas at $R_{trunc} \sim 150$ kpc around M87, but because the
total mass within the truncation radius appears to be already
dominated by cluster dark matter, a tidal truncation of the M87 mass
distribution may be difficult to see in X-ray observations.  However,
if M87 was currently tidally truncated by a tidal field with assumed
mass center towards M84/M86, we would expect to see some of the
tidally dissolved stars as PNe in our outer F7 fields at slightly
redshifted velocities with respect to the systemic velocity. Within
the limited statistics, we do not see any PNs with such velocities
outside a projected radius of 161 kpc, see \S~\ref{truncation}. This
suggests that if there has been a tidal truncation, it would have
occurred some time ago during the interaction with another mass
concentration.  The most likely candidate in the Virgo core may be
that around M84; at a relative velocity with respect to M87 of about
300 \kms, M84 could have travelled their current projected separation
in $\sim 1$ Gyr.

On the other hand, due to the dynamical youth of the Virgo cluster, it
is also possible that M87 has not been tidally affected yet, and is
more similar to an isolated massive elliptical galaxy. } As already
mentioned in \S~\ref{truncation}, the luminous halos around isolated
galaxies are expected from numerical simulations \citep{ans06} to
extend to the virial radius, i.e., to several hundred kpc, and well
beyond their traditional luminous radius. { Hence we now
consider the possible origins of the truncation of the M87 stellar
halo in the context of isolated galaxies. }

One possible explanation might lie in the fact that M87 is an old
galaxy with a massive nuclear black hole, which points to much
stronger AGN activity in the past than is apparent now. The feedback
from its AGN through the surrounding hot gas might at some redshift
$z_f$ have stopped the star formation in nearby satellite galaxies
{ through, e.g., ram pressure stripping}. When these satellites
later accreted onto the galaxy, they would have predominantly added
dark matter to the outer halo, so that the virial radius $R_V$ of M87
kept growing, but the luminous radius stalled at $R_V(z_f)$. On this
assumption, we can estimate $z_f$ from the redshift dependence of the
virial radius.

The X-ray observations show that the hot gas extends out to 300 kpc
\citep{nb95,mab02}. The derived integrated mass profile of the total
gravitating matter shows a change in slope at about 30 kpc
\citep{mab02}, and then increases linearly at large radii (see
Fig.~\ref{fig:M87_vcirc} above).  The mass distribution inferred from
the X-ray measurements thus provides evidence for two components: a
galaxy dark matter component and a cluster dark matter component. From
the modeling of \citet{nb95} and the rotation curve in
Fig.~\ref{fig:M87_vcirc} we can estimate the maximum circular velocity
generated by the M87 galaxy halo now to be $\vel_{\rm max,M87}\simeq
400\kms$.  Using the results of \citet{bullock+01}, this corresponds
to a present-day virial mass $M_V(z\!=\!0)\simeq
2.0\times10^{13}\Msol$ and virial radius $R_V(z\!=\!0)\simeq 470$ kpc,
several times larger than the truncation radius inferred from both the
PN number counts and the outer halo dynamics, $R_{trunc}\simeq 149$
kpc. For the same $\vel_{\rm max,M87}$, the virial radius of M87 would
have been $149$ kpc at redshift $z_f\simeq 2.9$, arguing that feedback
would need to have been effective quite early-on to explain the
observed truncation radius.

A second possible explanation would assume that the accretion of dark
matter and satellites onto M87 ceased with the collapse of the Virgo
cluster core. In the new potential after the collapse, the satellites
would both have been deflected from their nearly radial orbits with
respect to M87, and have significantly larger impact velocities than
previously, making accretion and merging with M87 suddenly less
likely. The total mass of M87 would thus not increase any further,
stalling at the virial mass at that redshift. Moreover, the rotation
curve in Fig.~\ref{fig:M87_vcirc} shows that with the on-going
collapse of the Virgo cluster a substantial cluster dark matter cusp
has since built up within the halo of M87. The likely effect of this
is an adiabatic contraction of the galaxy's luminous and dark halo.

The two-component mass model of \citet{nb95} (see equation
\ref{eq:mass}) for the present-day mass distribution predicts within
$R_{trunc} \simeq 149$ kpc, a galaxy mass of $M(R_{\rm trunc})_{\rm
  M87}\simeq 5.4\times 10^{12}\Msol$ and a cluster dark matter mass of
$M(R_{trunc})_{\rm Virgo}=1.2\times 10^{13}\Msol$, assuming a flat
rotation curve for the galactic contribution to the mass. The luminous
mass of M87, (4-5)$\times10^{11}\Msol$
\citep{cappellari+06}, is consistent with the estimated total galaxy
mass.  As an example, consider truncation of the accretion onto M87 by
the collapse of the Virgo core at redshift $z\!=\!0.5$. Using a lower
$\vel_{\rm c,mx,M87}\simeq 300\kms$ for the M87 halo before adiabatic
contraction and again the results of \citet{bullock+01}, the virial
mass and radius at that redshift become $M_V(z\!=\!0.5)\simeq
6.2\times10^{12}\Msol$ and virial radius $R_V(z\!=\!0.5)\simeq 300$
kpc.  An (over)estimate of the adiabatic contraction can be obtained
from angular momentum conservation at the outer radius, i.e., $G
M_V(z\!=\!0.5) R_V(z\!=\!0.5) = G [M(R_{trunc})_{\rm M87} +
M(R_{trunc})_{\rm Virgo}] R_{trunc}$, giving
$R_{trunc}/R_V(z\!=\!0.5)=0.3$.  This suggests that the observed
truncation radius could well be the relic of the virial radius from
the time when the cluster core collapsed.

{ It is clear that more data are needed to pursue this question
further. In particular, a larger number of PNs all around M87
would be very useful to set stronger constraints on the tidal
hypothesis.}

\section{The luminosity-specific PN number for the M87 halo and the
  ICL in Virgo\label{alphap}}

The physical quantity which ties a PN population to the luminosity of
its parent stars, is the luminosity-specific PN number $\alpha =
N_{PN} / L_{gal}$, where $N_{PN}$ is the number of all PNs in the
population\footnote{This is given by the integral over the whole eight
  magnitude range of the Planetary Nebula Luminosity Function (PNLF).}
and $L_{gal}$ is the bolometric luminosity of the parent stellar
population. Observations show that this quantity varies with the
$(B-V)$ color of the stellar continuum light \citep{hui93}, and simple
stellar population models predict that it is a function of the age and
metallicity of the parent stellar population \citep{bac06}.
Furthermore, within the framework of single stellar populations
models, the $\alpha$ parameter quantifies the average PN lifetime
$\tau_{PN}$ \citep{vill02,Ciardullo+05} via the relation $\alpha =
{\cal B} \tau_{PN}$, where ${\cal B}$ is the `specific evolutionary
flux' and is nearly constant \citep[see][for a detailed discussion]
{bac06}.  The PN samples in elliptical galaxies and the ICL
are all confined to the brightest 1 to 2.5 magnitudes of the PNLF.
Therefore we use the $\alpha_{2.5}$ parameter, defined in terms of
$N_{2.5}$, the number of PNs down to 2.5 magnitudes below the PNLF
cut-off: $\alpha_{2.5}$ equals about one tenth of $\alpha$ according
to the double exponential formula of \citet{Ciardullo+89} for the
PNLF.

We can use the number of photometrically detected PNs in the FCJ,
F7/F4 fields, and the luminosities of both the M87 halo and the ICL
populations sampled in the surveyed areas, to compute the
$\alpha_{2.5}$ values for these two components. Since the M87 halo and
the ICL coexist at the two field positions, we determine the fraction
of the PNs in the photometric sample bound to the M87 halo or in the
ICL according to the fraction of spectroscopically confirmed PNs
associated with each component in the LOSVD, i.e., with the narrow M87
peak or the nearly uniform velocity distribution for the ICL.

{\it Luminosity of the ICL} - We estimate the luminosities of the ICL
stellar population in our fields from the deep photometry of
\citet{mhfm05}. { Such surface brightness measurements for the
  diffuse light generally contain the cumulative contributions from
  extended galaxy halos, the true ICL, and from excess unresolved
  background galaxies above that adopted from the sky subtraction.
  Depending on the method of sky subtraction, the homogeneous part of
  the last component may be included in the sky measurement.  The
  photometry of \citet{mhfm05} shows a `plateau' of the surface
  brightness at a value of $\mu_V = 27.7$ half way between M87 and
  M86, where the F7/F4 fields are situated.  As we have seen in
  Sections~\ref{truncation}, \ref{sec:dynamics}, there is no
  contribution from the halo of M87 to the plateau. \citet{wcd07}
  estimate the surface brightness of background galaxies from their
  deep imaging survey with the Hubble Space Telescope's Advanced
  Camera for Surveys (ACS), in a small (intracluster) field within our
  field F4. These galaxies, which are resolved in the ACS data, would
  contribute a diffuse surface brightness of $\mu_V \simeq 28.6$ in
  ground-based data.  The sky subtraction procedure adopted by
  \citet{mhfm05} would have already subtracted this component if its
  surface brightness is similar in the edges of the mosaic where the
  sky was measured. In the following, we therefore use $\mu_V = 27.7$
  for the ICL surface brightness in this region, with a possible
  uncertainty of a few tenths of a magnitude due to a possible
  inhomogeneity of the background sources.}

{\it The $\alpha_{2.5}$ values and their implications} - In what
follows we make the assumption that the ICL surface brightness is
constant in the FCJ, F7/F4 fields at this value of $\mu_V \sim 27.7$ mag
arcsec$^{-2}$. In Table~\ref{tab:PNLTOT2} we give the corresponding
ICL luminosities and the number of spectroscopically confirmed PNs
from the ICL in these fields, $N_{spectr}$.  In
Table~\ref{tab:PNLTOT3} we give the number of PNs in the complete
photometric samples in the overlap area covered by the spectroscopic
follow-up, $N_{phot}$, and the fraction of PNs in the M87 halo and the
ICL according to their measured LOSVDs. Because of the small number
statistics, we compute $\alpha_{2.5}$ for the M87 halo in FCJ, and for
the ICL in F7, where their respective contributions are largest.

\begin{table}[h]
\caption{ICL PNs and sampled luminosities in the colored
regions of Fig.~\ref{area:red:green}.}
\begin{tabular}{lccc}
\hline 
Field   & N$_{spectr}$ &  Area        & L  \\  
        &              &  kpc$^2$     & L$_{V,\odot}$ \\
\\
FCJ red          &     2         &    4157       & $1.2 \times 10^9$ \\ 
F7 red + green   &     7         &    13258      & $3.8 \times 10^9$ \\
\hline\\
 
\end{tabular} \\
\label{tab:PNLTOT2}
\normalsize
\end{table}

\begin{table}[h]
\caption{Number of PNs in the photometric samples, for the M87 halo and 
the ICL.} 
\begin{tabular}{lccc}
\hline Field & N$_{phot}$ & N$_{phot}$& N$_{phot}$ \\ 
             &            & M87       & ICL         \\
\\ 
FCJ & 16$^a$ & 14 & 2 \\ 
F7  & 24$^b$ & 10 & 14 \\ 
\hline\\
 \end{tabular} \\
$^a$ Number of PNs above the photometric completeness limit from
Aguerri et al. (2005). $^b$ Number of PNs above the photometric
completeness from Table~\ref{tab:rates} in F7\_1 and F7\_2.
\label{tab:PNLTOT3}
\normalsize
\end{table}

The $N_{phot}$ of each field is scaled by a factor
\begin{equation}
\Delta=\frac{\int^{M^{*}+2.5}_{M^{*}} PNLF(m) dm}{\int^{m_{lim}}_{m^{*}}
PNLF(m) dm}
\end{equation}
where PNLF(m) is the analytic expression for the PNLF
\citep{Ciardullo+89}, $M^{*}$ and $m^{*}$ denote the absolute and
apparent magnitude of its bright cutoff, respectively, and $m_{lim}$
is the photometric $m_{5007}$ limiting magnitude in each field. This
scaling ensures that we account for all PNs within 2.5 mag of $M^{*}$.

The V-band luminosities in each field are converted to bolometric
luminosities according to
\begin{equation}
L_{bol} = L_{V,\odot}10^{-0.4(BC_V+0.07)}.
\end{equation}
According to \citet{bac06}, a value of $BC_V = -0.85$ mag can be taken
as a representative correction for all galaxy types within 10\%
uncertainty.

Finally, we obtain the bolometric luminosity-specific PN number
$\alpha_{2.5}$: For the M87 halo light at the FCJ position it is
$\alpha_{2.5,M87}= 3.1\times 10^{-9}$ PN L$_\odot^{-1}$, and for the
ICL at the F7 position it is $\alpha_{2.5,ICL}= 7.2 \times 10^{-9}$ PN
L$_\odot^{-1}$.  The values of $\alpha_{2.5}$ for different stellar
populations are well documented \citep{Ciardullo+05,bac06,coccato+09}:
$\alpha_{2.5}$ in the range $3-10\times 10^{-9}$ PN L$_\odot^{-1}$ are
observed for bright ellipticals and S0s. Both the $\alpha_{2.5}$
values obtained for the ICL and for the M87 halo stars are consistent
with those of old ($> 10$ Gyr) stellar populations.

{\it Uncertainties in the $\alpha_{2.5}$ values} - The luminosity
of the M87 halo is computed using Monte Carlo integration of the
S\'{e}rsic fit to the surface brightness from \citet{kormendy+08} in
the FCJ field, and the errors here are of the order of few percent.

The luminosity of the ICL is computed using $\mu_V = 27.7$ from
\citet{mhfm05}. We independently estimated the ICL surface brightness
by comparing the reprojected surface brightness profile of the M87
halo with the S\'{e}rsic fit of \citet{kormendy+08} in the F7 field.
{ This results in an azimuthally averaged ICL $\mu_V = 28.5$. While
$50\%$ fainter than the measurement of  \citet{mhfm05}, the two
values may be quite consistent when taking into account that the ICL
is observed mostly on the side of M87 towards M86/M84.}

Considering the uncertainties in the surface brightness for the M87
halo and ICL, and the statistical errors in the number of detected
PNs, the $\alpha_{2.5}$ values for M87 and ICL differ at the
$\sim2\sigma$ level. We speculate that $\alpha_{2.5,ICL}$ is a factor
2 larger than $\alpha_{2.5,M87}$ because of different metallicity
distributions in the ICL and the M87 halo, with a larger fraction of
metal-poor stars in the intracluster component, as shown by
\citet{wcd07}.

\section{ICL and the dynamics of the Virgo cluster core\label{VC}}

\subsection{ICPNs and dwarf spheroidals}

The LOSVDs in Figure~\ref{fig:vel_distrib} show the dynamical
components in the Virgo cluster core: the halo of M87, and the ICL
component traced by a broad PN velocity distribution. This component
covers the velocity range from 1300 \kms\ down to the
systemic velocity of M86 at -244 \kms. Overlaying the spatial
coordinates of the PNs on the deep image of the Virgo cluster core
(Figure~\ref{fig:vel_streams}; \citet{mhfm05}) we can easily see the
association of the PN components identified in the velocity - position
space, with the morphological components of the surface brightness
distribution in the Virgo core.  The M87 PNs are confined to still
relatively bright regions covered by the M87 halo, while the ICL PNs
are scattered across the whole region.

For comparison, we examine the phase space distribution of dwarf
elliptical galaxies in a region covering our target fields for the PNs
spectra, i.e., in a 1.5 degree diameter circle centred on the midpoint
of M87 and M86 (Figure~\ref{fig:vel_streams}). The aim is to search
for possible associations between our ICPNs and the positions and
velocities of the dwarf galaxies.  The top panel of
Figure~\ref{fig:vel_distrib} shows a histogram of the LOS velocities
for all dwarfs in the region marked in Figure~\ref{fig:vel_streams}.
The velocities form a flat, uniform distribution extending to larger
positive velocities than the ICPNs.  We ask whether any of the PNs
could be physically associated with the dwarf galaxies. There are only
two potential associations: one is between a PN with velocity 818\kms\
and a close-by dE at 791\kms. This dE has total blue apparent
magnitude 15.4, i.e., M$_B$=-15.48 (using the assumed distance) or
$L_B$=$1.4\times10^8\Lsol$. According to \citet{Ciardullo+05},
galaxies fainter than M$_B\sim-19$ and bluer than $V-I<1.1$ produce
about one \Oiii -bright PN in every $4\times10^8\Lsol$ . It is
therefore unlikely that the dwarf galaxy in question produced the PN
detected here (as the expected number is 0.35) although it cannot be
ruled out completely.  On the other hand, the second association may
well be genuine: this is of a PN with velocity +28\kms\ and a close-by
Sb spiral galaxy to its west, which has velocity 30\kms\ and
$B_{tot}$=10.91 (and therefore is capable of producing between 85 and
2000 PNs depending on its age, see Buzzoni et al. 2006\nocite{bac06}).

Although the majority of the ICPNs do not appear to be
physically associated with the dwarf galaxies (i.e., the PNs are
unlikely to originate in the dwarfs), their distribution in velocity
space is at least partially similar, indicating that they follow
similar dynamics.

\subsection{Dynamical status of the Virgo core}

The velocity distribution of dwarf spheroidals (dE+dS0) in a $2^\circ$
radius circular region centred on M87 is very flat and broad, with the
peak of the distribution at 1300 \kms\ and a long tail of negative
velocities \citep{bpt93}.  The LOSVD of the ICPNs now confirms that
this asymmetry is also present in the very center of the Virgo core,
in a region of $1^\circ$ diameter. { Fig.~\ref{fig:phasespace}
  shows that velocities near the systemic velocity of M86 are seen to
  about half-way from M86 to M87.}

The asymmetry and skewness of the LOSVD may arise from the merging of
subclusters along the LOS as described by \citet{sb93}. In their
simulations of two merging clusters of unequal mass, the LOSVD is
found to be highly asymmetric with a long tail on one side and a
cut-off on the other side, shortly ($\sim 10^9$ yrs) before the
subclusters merge. 

The observed LOSVDs of the PNs, GCs \citep{cmh+01}, and (dE+dS0) in
the Virgo core may therefore be interpreted as additional evidence
that the two massive subclusters in the Virgo core associated with the
giant ellipticals M87 and M86 are currently falling towards each other
- more or less along the LOS, with M87 falling backwards from the
front and M86 forwards from the back - and will eventually merge,
i.e. the entire core of the Virgo cluster must then be out of virial
equilibrium and dynamically evolving.

The distribution of the brightest galaxies in Virgo also favors a
recent and on-going assembly: West \& Blakeslee (2000)\nocite{wb00}
found that Virgo's brightest elliptical galaxies tend to be aligned
along the principal axis of the cluster (which is inclined by only
about 10-15 degrees to the line of sight) and which on larger scales
connects Virgo to the rich cluster Abell 1367. This work suggests that
the formation of the cluster is driven by infall along this filament.

Do the halos of M87 and M86 already touch each other, or are they just
before their close pass?  PNLF distances \citep{jcf90} and
ground-based surface brightness fluctuation distances \citep{tdb01}
indicate that M86 is behind M87 by just under $\sim$0.15
magnitudes. The globular cluster LF turnover also suggests that M86 is
likely 0.1 to 0.2 magnitudes more distant than the main body of Virgo
\citep{kw01}.  However, the most recent surface brightness fluctuation
measurements by Mei et al. (2007)\nocite{mbc+07} find that M87 and M86
are only at very slightly different distances. Within the errors, the
distance moduli (M87: $31.18\pm0.07$, M86: $31.13\pm0.07$) are
consistent with being either at the same distance or separated by 1-2
Mpc. Unfortunately the evidence from the relative distances of
M87/M86 is not conclusive at this stage.

\subsection{Implications for the formation of the ICL}

The observational facts concerning the ICL in the Virgo cluster core
are: 
\begin{enumerate}
\item The LOSVD of the ICPNs is not symetrically distributed around
  the systemic velocity of M87. Those between M87 and M86 are mostly 
  at `bluer' velocities, i.e., $ < 800$ \kms. { ``Red'' velocities are
  only seen in the field $0.^\circ8$ north of the line connecting
  M87 with M86; see Fig.~\ref{fig:vel_streams}.}
\item While the dwarf spheroidals' (dE+dS0) LOSVD in the region marked
  in Figure~\ref{fig:vel_streams} extend into `red' velocities, up to
  2500 \kms, { ICPNs with velocities greater than 1800 \kms are seen
  only at its northern perimeter, while those in the region between
  M87 and M86 are confined to $<800$ \kms (see
  Figs.~\ref{fig:vel_distrib}, \ref{fig:vel_streams},
  \ref{fig:phasespace}).} This is not a consequence of the filter used
  in the photometric selection of these objects, which still has a
  transmission of 50\% to [OIII]$\lambda$5007 at $\sim$2275\kms\
  \citep{fcjd03}.  \footnote{There may be a different, small selection
    effect due to the finite limiting magnitude of the photometric PN
    survey, as the ICPN that we detect are slightly biased towards
    objects on the near side of the cluster core.}
\item The morphology of the ICL between M87 and M86 is `diffuse'; 
  it is mostly not in tidal tails or streams \citep{mhfm05}.
\item The measured $\alpha_{2.5}$ parameter for the ICL is in the
  observed range for old stellar populations.
\item { The metallicity distribution of the RGB stars associated
    with the ICL in the \citet{wcd07} field is broad, with a peak at
    about 0.1 solar, and the best model of \citet{wcd07} indicates an
    old stellar population ($> 10$Gyr).}
\end{enumerate}

Point 1) indicates that the ICL did not have the time to phase mix
yet, supporting the idea that much of this diffuse component is
falling towards M87 together with the M86 group. Points 2) and 3)
argue against the origin of this ICL from current harrassment of dwarf
irregulars by the cluster potential \citep{mkl96}, firstly because
none of the ICPNs between M87 and M86 are found at velocities redder
than 800 \kms, while we see nearby dwarf galaxies in the $ 1800 <
\vel_{LOS} < 2500$ \kms\ range, and secondly because the diffuse
morphology of the ICL suggests that it has not recently been dissolved
from the dwarfs, but had time to phase mix at earlier times in the
sub-component of the Virgo cluster which is now falling towards M87.
Points 4) and 5) indicate that the parent stellar population of the
ICPNs is an old population, and { point 5) makes dwarf spheroidals
  unlikely progenitors (recall that the dSph analyzed by
  \citet{durrell07} had a narrow metallicity distribution at
  [Fe/H]$\simeq-2$). Dwarf elliptical galaxies have a wider
  metallicity distribution, and they could be disrupted during
  passages through the high-density regions around M87. Some of the
  observed ICPNs could trace stars on a stream from such a disruption
  event (particularly, those at $\vel_{\rm LOS}\sim 800$ \kms), but
  those at low velocities in Fig.~\ref{fig:phasespace} are falling in
  with M86, so have not crossed the dense regions yet. Point 3) then
  implies that these must have been part of the diffuse halo of the
  M86 group. }

We conclude that we have found observational evidence in the Virgo
core for the mechanism described by \citet{rmm06}: we observe the
diffuse component `pre-processed' in the M86 sub-group, which is or
has been gravitationally unbound from M86 as this substructure is
being accreted by M87. The idea that the diffuse light is being
stripped from the M86 sub-group is consistent with the observed highly
skewed LOSVD and with the predictions from the simulations of
\citet{sb93}. Note that the light in the M86 subgroup is tidally
stripped by the more massive M87 component, while these two
sub-structures merge along the LOS; we do not see a diffuse ICL with a
broad velocity component redwards of the systemic velocity of M87,
because it has not yet been formed.

This scenario is also consistent with the simulations of
\citet{mgga+07}. Their statistical analysis of the diffuse star
particles in a hydrodynamical cosmological simulation indicates that
most of the ICL is associated with the merging tree of the brightest
cluster galaxy, and about 80\% of the ICL is liberated shortly before,
during and shortly after major mergers of massive galaxies. The
results from \citet{mgga+07} imply that the main contribution to the
ICL comes from merging in earlier sub-units whose merger remnants
later merge with the final cD galaxy.  Similarly, \citet{rmm06}
predict that violent merging events quickly add ICL, and without or
between these events, the ICL fraction rises only slowly.  Once the
M86 subgroup has finally merged with M87, this will have created the
most massive galaxy in the then Virgo cluster, and the ICL in the
future Virgo core will indeed have originated mainly from the
progenitors associated with its merging tree.

\section{Summary and Conclusions\label{last}}

{ 

Using high resolution multi-object spectroscopy with FLAMES/MEDUSA on
the VLT we confirm a further 12 PNs in the Virgo cluster, located
between $130$ and $250$ kpc from the center of M87, and obtain their
radial velocities. For most of these objects we also detect the second
line \Oiii$\lambda$ 4959\AA. These PNs trace the kinematics of diffuse
light in Virgo, at typical surface brightness of $\mu_V=27.5$.

The phase-space distribution for the new sample of PN velocities
combined with earlier measurements at $60$ and $150$ kpc from
M87 illustrates the hierarchical nature of structure formation. One
group of PNs has an unrelaxed distribution of velocities with a range
characteristic for the still assembling Virgo cluster core, while the
second group has a narrow velocity distribution which traces the
bound, cold outer halo of M87. We summarize our results for these two
groups in turn.

\subsection{Dynamical status of the Virgo cluster and origin of the
ICL\label{originICL}}

Seven of the newly confirmed PNs are genuine intracluster PNs in the
Virgo core, not bound to M87.  Their spatial and velocity distribution
indicates that we are witnessing the gravitational stripping of the
diffuse light component around the M86 group, as this sub-structure is
being accreted by the more massive M87. We do not see a diffuse ICL
with a broad velocity distribution including red-shifted velocities
around M87, because it presumably has not been formed yet.

On the basis of the LOSVDs of ICPNs and galaxies in the Virgo core, we
surmise that M87 and M86 are falling towards each other nearly along
the line of sight, and that we may be observing them in the phase just
before the first close pass.  We thus conclude that the heart of the
Virgo cluster is still far from equilibrium.

Finally, the $\alpha_{2.5}$ values determined for the ICL indicate an
old stellar population. This is consistent with the analysis by
\citet{wcd07} of the colour-magnitude diagram (CMD) for ICL red giant
stars, which showed an old stellar population ($\gta10$ Gyr) with a
large spread of metallicities. Differently, the CMD for a nearby dwarf
spheroidal galaxy indicates a similarly old, but metal-poor stellar
population \citep{durrell07}.  Together with the observed $\vel_{LOS}$
distribution of the ICPNs, these results suggest that at least some of
the ICL in Virgo originates from stars unbound from the brightest and
most massive galaxies.

}

\subsection{The M87 halo}
The other five of the newly confirmed PNs are associated with the
bound halo of M87, at a mean projected radius $R=144$ kpc from
the centre of the galaxy. These PNs have velocities close to the
systemic velocity of M87, with a small dispersion, and are furthermore
segregated spatially from the rest of the (intracluster) PNs, as shown
in Figs.~\ref{fig:vel_streams} and \ref{fig:phasespace}.

The LOSVDs of the M87 PNs both in the new fields and in the FCJ field
of A04 are consistent with no rotation of the outer halo around the
photometric minor axis, outside $R = 15$ kpc. We cannot test whether
the halo is rotating around the major axis.  The rms velocity
dispersion for the 5 M87 PNs at 144 kpc is $78 \pm 25$ \kms, much
smaller than the central velocity dispersion. Together with the
results of A04, this indicates that the M87 halo becomes `dynamically'
cold beyond 50 kpc radius.

The PNs around the systemic velocity of M87 are confined to radii
$R\le161$ kpc.  The absence of M87 PNs at larger radii with respect to
the extrapolated S\'{e}rsic fit to the surface brightness profile from
\citet{kormendy+08}, despite being based on small numbers, is
significant at the $\sim 2\sigma$ level. This suggests that the edge
of M87 has been detected, and it occurs at quite a large average
ellipse radius - $R_{trunc} = 149$ kpc.

We have tested the hypothesis of a truncated stellar halo dynamically,
using the observed stellar kinematics of the M87 halo. Combining all
the velocity dispersion data available in the literature with our new
M87 $\sigma_{PN}$ data, we have solved the spherical Jeans equation
assuming a total gravitational potential as traced by the X-ray
emission. Within this framework, the Jeans model is able to reproduce
the `cold' PN velocity dispersion in the outer halo of M87 only if (i)
the orbital structure in the outer halo becomes highly radially
anisotropic, with $\beta > 0.4$ at $r > 10$ kpc, {\sl and} (ii) the
intrinsic luminous density is truncated.  This dynamical argument can
be circumvented only if the stellar distribution were strongly
flattened along the line-of-sight in the surveyed fields. At these
radii this flattening would be local and imply that the spheroidal
stellar halo would end at even smaller radii.

The evidence for the truncation of the luminous halo of M87 thus comes
from both the spatial distribution of the PNs with velocities near the
systemic velocity of M87 (Fig.~\ref{fig:phasespace}), and from the
small velocity dispersions in both the A04 field and in the new fields
near the outer edge. 

The reason for the truncation is not obvious; we discuss possible
mechanisms in Section~\ref{origint}. Differently from some dense
clusters where lensing analysis indicates that galaxies outside the
cluster center are tidally truncated by the dark matter cusp of the
cluster, M87 in Virgo is located at the center of the deepest
potential well traced by the X-ray isophotes. We also do not see
unbound PNs with velocities near 1300 km/s further out from M87.  This
suggests that if there has been a tidal truncation, it would have
occurred some time ago during the interaction with another mass
concentration, such as around the other massive galaxy in the Virgo
core, M84. Alternatively, due to the dynamical youth of the Virgo
cluster, it is also possible that M87 has not been tidally disturbed
yet, and is more similar to an isolated massive elliptical galaxy and
should thus still be accreting matter \citep{ans06}. In this case
possible explanations for the truncation could be early AGN feedback
effects that indirectly truncate star formation in accreting
satellites, or adiabatic contraction of the M87 halo due to cluster
dark matter collapsing onto the galaxy.

The existing data cannot discriminate between these scenarios.  The
next step in this project is therefore to obtain a sample of at least
a few hundred measured PN velocities covering the whole M87 halo. This is
required to verify that the velocity dispersion decreases everywhere
around the galaxy, and to obtain statistically better constraints on
phase-space structures in the surrounding ICL, including possible stars
tidally dissolved from M87. With a homogenous imaging and
spectroscopic PN survey covering the whole halo of M87 out to 40
arcmin we will be able to accurately measure the rotation, radial
anisotropy of the orbits, and truncation of the outer halo of M87.  

\begin{acknowledgements}
  We wish to thank Nando Patat for carrying out the observations in
  service mode, and Marina Rejkuba and Sandro Villanova for advice in
  using the GIRAFFE pipeline.  We thank Ken Sembach for providing his
  velocity dispersion data in digital format, John Kormendy for giving
  us the surface brightness profile data for M87 and the parameters of
  the best S\'{e}rsic fit before publication, and Ralf Bender, James
  Binney and Karl Gebhardt for useful discussions. PD was supported by
  the DFG Cluster of Excellence.
\end{acknowledgements}

\bibliographystyle{aa}
\bibliography{1532}

\end{document}